\documentclass[a4paper,11pt]{article}
\pdfoutput=1 
\usepackage{jheppub}

\usepackage[utf8]{inputenc}

\usepackage{amssymb,amsmath,amsfonts}
\usepackage{mathtools}
\usepackage{mathrsfs}
\usepackage{bbm}
\usepackage{slashed}
\usepackage{nicefrac}
\usepackage{graphicx}
\usepackage{caption}
\usepackage{subcaption}
\usepackage{esint}

\usepackage{graphicx}
\usepackage[dvipsnames]{xcolor}
\usepackage{array}

\usepackage{simplewick}

\usepackage{hyperref}
\usepackage{xparse}
\usepackage{xspace}

\usepackage{tikz}
\usetikzlibrary{decorations.pathmorphing}
\usetikzlibrary{automata,positioning}

\usepackage{cancel}
\usepackage[normalem]{ulem}

\usepackage{xifthen}
\usepackage{dsfont}
\usepackage[titletoc]{appendix}
\usepackage{booktabs}
\usepackage{units}

\newcommand{\gettitle}{}
\hypersetup{linkcolor=black
	colorlinks,
	linkcolor={red!75!black},
	citecolor={blue!75!black},
	urlcolor={blue!75!black},
	pdftitle={\gettitle},
	pdfauthor={Barata, Mehtar-Tani},
	pdfkeywords={Perturbative QCD} {Jet quenching},
	bookmarksopen=true,
	bookmarksopenlevel=2,
	bookmarksnumbered=true
}
\makeatletter
\newcommand\makebig[2]{%
  \@xp\newcommand\@xp*\csname#1\endcsname{\bBigg@{#2}}%
  \@xp\newcommand\@xp*\csname#1l\endcsname{\@xp\mathopen\csname#1\endcsname}%
  \@xp\newcommand\@xp*\csname#1r\endcsname{\@xp\mathclose\csname#1\endcsname}%
}
\makeatother
\makebig{biggg} {1.3}

\def\bs{\boldsymbol} 

\def\bdel{\bs\partial}

\newcommand{\eqn}[1]{Eq.~\eqref{#1}}

\long\def\comment#1{ }

\newcommand{\nn}{\nonumber\\ }

\def\be{\begin{eqnarray*}}
\def\ee{\end{eqnarray*}}
\def\beq{\begin{eqnarray}}
\def\eeq{\end{eqnarray}}
\newcommand{\bea}{\beq \begin{aligned}}
\newcommand{\eea}{\end{aligned}\eeq}

\def\k{{\boldsymbol k}}

\def\q{{\boldsymbol q}}

\def\x{{\boldsymbol x}}
\def\y{{\boldsymbol y}}
\def\z{{\boldsymbol z}}

\def\0{{\boldsymbol 0}}

\def\k{{\boldsymbol k}}

\def\x{{\boldsymbol x}}
\def\y{{\boldsymbol y}}

\def\z{{\boldsymbol z}}

\def\sst{\scriptscriptstyle}
\def\rme{{\rm e}}
\def\rmd{{\rm d}}

\def\and{ \quad\text{and}\quad}

\def\Re{\text{Re}}
\def\Im{\text{Im}}

\def\mfp{\text{mfp}}

\def\abar{{\rm \bar\alpha}}

\def\cK{{\cal K}}

\def\omegaBH{\omega_{\rm \sst BH}}
\def\BH{{\rm BH}}

\def\abar{\bar\alpha}

\def\abar{\bar\alpha}

\begin{document}

\title{Improved opacity expansion at NNLO for medium induced gluon radiation}

\author[a,b]{Jo\~ao Barata}
\author[a,c]{Yacine Mehtar-Tani}

\affiliation[a]{Physics Department, Brookhaven National Laboratory, Upton, NY 11973, USA}
\affiliation[b]{Instituto Galego de F\'{i}sica de Altas Enerx\'{i}as (IGFAE), Universidade de Santiago de Compostela,
E-15782 Galicia, Spain}
\affiliation[c]{RIKEN BNL Research Center, Brookhaven National Laboratory, Upton, NY 11973, USA}

\emailAdd{joao.barata@cern.ch}
\emailAdd{mehtartani@bnl.gov}

\date{\today}
\abstract{When an energetic parton propagates in a hot and dense QCD medium it loses energy by elastic scatterings or by medium-induced gluon radiation. The gluon radiation spectrum is suppressed at high frequency due to the LPM effect and encompasses two regimes that are known analytically: at high frequencies $\omega >\omega_c = \hat q L^2$, where $\hat q $ is the jet quenching transport coefficient and $L$ the length of the medium, the spectrum is dominated by a single hard scattering, whereas the regime  $\omega <\omega_c$ is dominated by multiple low momentum transfers. In this paper, we extend a recent approach (dubbed the Improved Opacity Expansion (IOE)), which allows an analytic (and systematic) treatment beyond the multiple soft scattering approximation, matching this result with the single hard emission spectrum. We calculate in particular the NNLO correction analytically and numerically and show that it is strongly suppressed compared to the NLO indicating a fast convergence of the IOE scheme and thus, we conclude that it is sufficient to truncate the series at NLO.  We also propose a prescription to compare the GW and the HTL potentials and relate their parameters for future phenomenological works. 

}

\date{\today}
\maketitle
\flushbottom


\section{Introduction}\label{sec:intro}

The strong modification of jet observables in Heavy Ion collisions (measured both at RHIC\cite{RHIC1,RHIC2} and the LHC\cite{LHC1,LHC2,LHC3}) when compared to proton-proton events, provides one of the key observations of the formation of the Quark Gluon Plasma (QGP) in such events. 
The continuous interactions between these hard probes and the dense QCD plasma induce a cascade of gluons, which inevitably modify the jet's properties (see \cite{Review_Blaizot_Yacine,Review_YKG} for recent reviews on the topic). As a consequence, for extracting the QGP properties from the experimental study of jets, an accurate and complete understanding of the medium induced radiation spectrum is critical. \par
One of the first (and most crucial) theoretical steps towards this goal consisted on the study of the emission spectrum  of a single parton embedded in a QCD medium. In the regime where the medium is sufficiently large 
such that the parton may interact with several scattering centers in the plasma, the medium induced spectrum admits a full analytic treatment, captured by the Baier-Dokshitzer-Mueller-Peigné-Schiff-Zakharov (BDMPS-Z \footnote{As was later shown \cite{Wiedemann}, this formalism also includes the regime of single hard scattering explored in the Gyulassy-Levai-Vitev (GLV) framework \cite{GLV}.}) formalism \cite{BDMPS1,BDMPS2,BDMPS3,BDMPS4,BDMPS5}. The region of validity for the BDMPS-Z formalism is bounded from below by the single (low energy) scattering limit (Bethe-Heitler limit), where the quantum mechanical formation time of the radiated gluon is of the order of the in-medium mean free path, $ t_f\equiv \omega/k_\perp^2 \sim \ell_\mfp$ which is assumed to be much smaller than the medium length $L$, and thus the gluon is emitted incoherently  by individual scattering centers. On the opposite end, the gluon formation time is bounded from above by the medium length $L$. In this regime, multiple soft scattering may act coherently as a single scattering center during $t_f$. Hence, the effective number of scattering centers is much smaller than the actual number of scattering centers, i.e., $N_{\rm eff} \sim L/t_f \ll L/\ell_\mfp \equiv N_{\rm scatt}$. The transverse momentum accumulated during $t_f$ via diffusion, $k_\perp^2 \sim \hat q t_f$, where $\hat q$ is the corresponding transport coefficient. This allows us to solve for the formation time 

\beq 
t_f =\frac{\omega}{k_{\perp}^2}\sim \frac{\omega}{\hat q \,t_f} = \sqrt{\frac{\omega}{\hat q }} \ . 
\eeq
Hence, the radiative spectrum is suppressed as $N_{\rm{ eff}} \sim \omega ^{-1/2}$ for $\omega_\BH = \hat q \ell_\mfp^2 \ll \omega \ll \omega_c = \hat q L^2 $. This is the QCD analog of the Laudau-Pomerantchuk-Migdal (LPM) effect \cite{LPM1,LPM2}. For formation times larger than $L$ a maximum LPM suppression is achieved. \par
The above parametric analysis is valid so long as one can neglect large momentum transfers from the medium to justify the application of the diffusion approximation. However, due to the large Coulomb tail in the elastic cross section the medium transport parameter $\hat{q}$ will depend logarithmically on the transverse size of the radiated gluon. In the BDMPS-Z approximation, one assumes that $\hat q$ is roughly constant invoking the slow variation of the Coulomb Logarithm. This makes the problem analytically tractable, but fails to capture the correct scaling in the region of phase space where the dominant contribution comes from single hard scattering (which is correctly captured by the GLV approach).\par 
Until recently, an analytic approach which was able to connect the BDMPS-Z and GLV regimes into a single framework was not known, although several numerical based approaches were able to solve the problem exactly \cite{numerical1,numerical2,numerical3}. In previous papers, one of us introduced a systematical way of taking into account the hard $p_T$ tail encapsulated in the medium scattering potential \cite{Paper1}. This approach was latter extended to also take into account the full scattering potential \cite{Paper2}, and was shown to correctly capture the features of both regimes\footnote{More recently, another numerical approach \cite{CarlotaFabioLiliana}, similar to that proposed in \cite{numerical1}, was able to resum the contribution from multiple scatterings with the full potential by providing a numerical recipe to solve the associated transport equation.}. In this paper, we will refer to this approach as the Improved Opacity Expansion (IOE).\par
We extend the work presented in \cite{Paper1} by computing the next order contribution to the integrated medium induced emission spectrum in the IOE approach. We study the NNLO term (i.e. we allow for the possibility of two hard scattering centers) in the IOE, showing that this term gives a small contribution to the full spectrum (when compared with the LO (BDMPS-Z) and NLO terms), ensuring that the series expansion is under control. \par 

We show, in particular that in contrast to the plain opacity expansion where high orders are suppressed by inverse powers of $\omega$, which is indeed the case for $\omega >\omega_c$, in the regime $\omega< \omega_c$ higher orders are only suppressed logarithmically and the leading order power scaling $\omega^{-1/2}$ extends to all orders. 
As a result, the full spectrum in this regime can be expressed in the leading order form with an effective transport coefficient that can be calculated order by order in the IOE scheme, that is, 
\beq
\omega\frac{\rmd I}{\rmd \omega\rmd L} = \abar \sqrt{\frac{\hat q _{\rm eff}(Q_c)}{\omega}} \, ,
\eeq
where the effective transport coefficient is calculated to NNLO in the IOE
\beq
\hat{q}_{\rm eff}(Q_c) = \hat{q}_0 \log\left(\frac{Q_c^2}{\mu^{\star2}}\right)\left[1+\frac{1.016}{\log\left(\frac{Q_c^2}{\mu^{\star2}}\right)}+\frac{0.316}{\log^2\left(\frac{Q_c^2}{\mu^{\star2}}\right)} +{\cal O}\left(\log^{-3}\left(\frac{Q_c^2}{\mu^{\star2}}\right)\right)\right]\, ,
\eeq
evaluated at the scale 
\beq 
Q^2_c = \sqrt{ \hat q_0\, \omega \log\left(\frac{Q_c^2}{\mu^{\star2}} \right)}\, ,
\eeq
where the IR cut-off's, that are fully fixed at leading logarithm accuracy, read
\beq
\mu^{\star2} =    \begin{dcases}
      \,   \frac{\mu^2}{4}\, \rme^{-1+2 \gamma_E} & \text{for GW model}\\
      \,  \frac{m_{\rm D}^2}{4}\, \rme^{-2+2 \gamma_E} & \text{for HTL model}\,,\\
    \end{dcases} 
\eeq
and $\hat q_0$ is given by \eqn{eq:qhat_0} and \eqn{eq:qhat_HTL} for the GW and the HTL models respectively.

The present manuscript is divided as follows. Section \ref{sec:medium_induced_spectrum} and subsections therein review the work presented in \cite{Paper1} and introduce a general form for the IOE expansion. In section \ref{sec:The_Next_to_Next_to_Leading_order} we study the NNLO order term in the IOE. Finally we discuss and summarize our findings in section \ref{sec:Discussion_of_results}. Complementary numerical work is shown throughout the paper.

\section{Medium-induced gluon spectrum}\label{sec:medium_induced_spectrum}
The general form for the integrated medium-induced gluon spectrum off a high energy parton (in color representation $R$) is given by \cite{Paper1,BDMPS5}
\begin{equation}\label{eq:spectrum_general}
\omega \frac{\rmd I}{\rmd\omega}=\frac{\alpha_sC_R}{\omega^2}\int_0^{\infty}dt_2\int_0^{t_2}dt_1 \ \bdel_\x \cdot \bdel_\y \left[\cK(\x, t_2|\y,t_1)-\cK_0(\x, t_2|\y,t_1)\right]_{\x=\y=0} \, ,
\end{equation}
where $\omega$ is the gluon frequency (assumed to be much softer than the emitting parton $E\gg \omega$) and the second term inside the brackets subtracts the vacuum like contributions. The Green's functions $\cK$ and $\cK_0$ are solutions to a 2-dimensional Schr\"{o}dinger equation in the transverse plane, and obey
\begin{equation}\label{eq:cK_Sch}
\left[i\partial_t+\frac{\bdel^2}{2\omega^2}+iv(\x)\right]\cK(\x,t_2|\y,t_1)=i\delta(\x-\y)\delta(t_2-t_1) \, ,    
\end{equation}
where $v(\x)$ is the potential defined by the in-medium (elastic) scattering cross section. $\cK_0$ obeys a similar equation with $v=0$.\par

\subsection{The HTL and Gyulassy-Wang potentials}

Typically, the in-medium scattering cross section can either be obtained from Hard-Thermal-Loop (HTL) theory or from the Gyulassy-Wang model (GW model). Since in this paper we only focus on the large $k_\perp$ tail corrections, to leading logarithmic accuracy the model choice is irrelevant. However, even at leading order both models differ by finite terms. As such, the choice for the medium parameters, in each model, has to take this into account by providing a map between the different model parameters and the set of physical parameters. Therefore, before presenting the Improved Opacity Expansion in order to solve \eqref{eq:spectrum_general}, we compute the potential $v$ entering \eqref{eq:cK_Sch} to leading order accuracy in both the GW and HTL models. As we will show, this will allow not only to have the full leading term of the potential in both models, but also provides a map between each model.\par
The elastic cross section in the GW model corresponds to an Yukawa interaction and reads \begin{equation}\label{eq:Xsec_GW} 
\left(\frac{\rmd^2\sigma}{\rmd^2\q}\right)^{\rm GW}=\frac{g^4n(t)}{(\q^2+\mu^2)^2}\, ,
\end{equation} 
where $n$ is the density of scattering centers in the medium, $\mu$ is an infrared cut-off related to the Debye mass in the plasma $m_{\rm D}$ and $g$ is the QCD coupling constant. In general, $n$ is a function of time, which we will overlook for the moment. Then the potential $v$ appearing in \eqref{eq:cK_Sch} reads (see Appendix \ref{app:integrals} for derivation)

\begin{equation}\label{eq:v_GW}
\begin{split}
v(\x,t)^{\rm GW}&=C_A\int_\q   \left( \frac{d^2\sigma}{d^2\q}\right)^{\rm GW} (1-e^{i\q\cdot\x})
\\&=\frac{\hat{q}_0}{\mu^2}\Big[1-\mu |\x| K_1(\mu |\x|)\Big] \, ,
\end{split}
\end{equation}
where we have explicitly introduced the color charge $C_A$ directly into the potential and $K_1$ is the modified Bessel function of the second kind of order 1. Following \cite{Paper1} we have introduced the transport coefficient stripped of any logarithm $\hat{q}_0$
\begin{equation}\label{eq:qhat_0}
\hat{q}_0(t)\equiv 4 \pi \alpha_s^2 C_A n(t)  \, , 
\end{equation}
with $\alpha_s=g^2/ (4\pi)$ and $\gamma_E\approx 0.577$ is the Euler-Mascheroni constant. Expanding \eqref{eq:v_GW} to leading accuracy we obtain
\begin{equation}
\begin{split}
v(\x,t)^{\rm GW}&= \frac{\hat{q}_0}{4}\x^2\log\left(\frac{4 \rme^{1-2\gamma_E} }{\x^2 \mu^2}\right)+\mathcal{O}(\x^4 \mu^2)\equiv \frac{\hat{q}_0}{4}\x^2\log\left(\frac{1}{\x^2\mu^{\star2}}\right)  \, ,
\end{split} 
\end{equation}
where we have introduced the physical scale $\mu^\star$ related to the GW screening mass $\mu \equiv\mu_{\rm GW}$ by $\mu^{\star2}\approx \mu_{\rm GW}^2 \ 0.29$. The subdominant term is suppressed by a power of  $\x^4 \mu^2$.\par 
The HTL formalism \cite{Aurenche:2002pd} predicts an elastic cross section of the form
\begin{equation}\label{eq:Xsec_HTL}
\left(\frac{\rmd^2\sigma}{\rmd^2\q}\right)^{\rm HTL}=\frac{g^2 m^2_{\rm D}T}{\q^2(\q^2+m_{\rm D}^2)}\, ,
\end{equation} 
where $m_{\rm{D}}^2=(1+\frac{n_{\rm{f}}}{6})g^2T^2$ is the QCD Debye mass (squared), $T$ is the QCD plasma temperature and $n_{\rm f}$ is the number of light quark degrees of freedom. For an equilibrated system it is well known that $n\sim T^3$ and the HTL and GW give the same result when the Debye mass is only taken into account as the infrared cut-off. \par
\begin{figure}[h!]
    \centering
    \includegraphics[scale=.6]{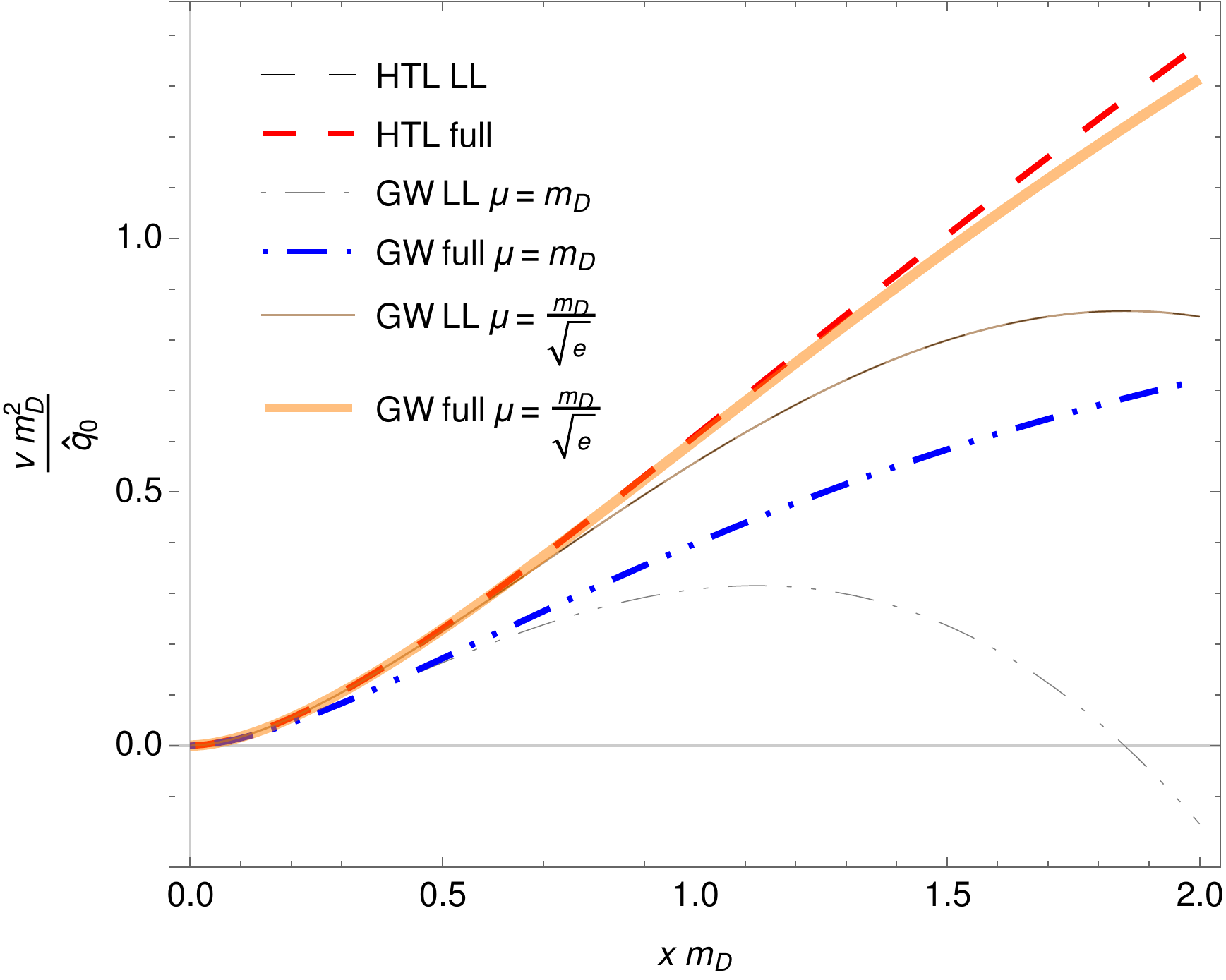}
    \caption{Plot of the potential $v$ for the HTL and GW models. Here we have normalized the potential to $\frac{m_{\rm D}^2}{\hat{q}_0}$ and the dipole size $|\x|$ is given in units of the Debye mass. The dashed curves correspond to the HTL model, the dash-dotted lines give the GW model potential when $\mu_{\rm GW}=m_{\rm D}$ and the full lines correspond to the GW model solution in the Leading Logarithmic (LL) approximation (full thin curve) and for the full potential (full grosser line) when one makes use of the matching proposed in \eqref{eq:GW_HTL_match}. The LL curves for both the HTL and GW model show that this approximation breaks down when $|\x|\sim \frac{1}{m_{\rm D}}$, as expected.  }
    \label{fig:GW_vs_HTL}
\end{figure}

 
Solving for $v^{\rm HTL}$ analytically is similar  to $v^{\rm GW}$ (see Appendix \ref{app:integrals}). We find
\beq\label{eq:v-HTL}
v(\x,t)^{\rm HTL}&=&\frac{g^2C_Am_{\rm D}^2 T}{2\pi}\int_0^\infty dq \ \frac{q}{q^2\left(q^2+m_{\rm D}^2\right)}\left(1-J_0(q|\x|)\right)    \, \nn
&=&   \frac{ 2\hat q_0 }{m^2_{\rm D}} \left[ K_0(m_{\rm D}|\x|) + \log(m_{\rm D}|\x|)-\log(2) +\gamma_E \right]\,,
\eeq
where now $\hat q_0$ is given by 
\beq\label{eq:qhat_HTL}
 \hat q_0 \equiv \alpha_s C_A m^2_{\rm D} T. 
\eeq
Expanding \eqn{eq:v-HTL} for small transverse size $\x$ we obtain 
\begin{equation}
v(\x,t)^{\rm HTL}=\frac{\hat{q}_0}{4}\log\left(\frac{4 \rme^{2-2 \gamma_E}}{m^2_{\rm D} \x^2 }\right)\equiv\frac{\hat{q}_0}{4}\log\left(\frac{1}{\mu^{\star 2}\x^2}\right) \, .
\end{equation}
This provides a complete and consistent (leading order) map between the set of parameters used in the GW model and the set of parameters coming from the HTL framework. In particular if we want to match the HTL and GW models to logarithmic accuracy we should have
\begin{equation}\label{eq:GW_HTL_match}
  m^2_{\rm D} =\rme\,  \mu^2_{\rm GW}   \,.
\end{equation}
In conclusion, we have shown that the leading logarithmic behavior for the potential entering equation \eqref{eq:cK_Sch} is fully captured in the GW model and HTL approach by defining a physical screening mass $\mu^\star$ and a map between the medium model parameters and this scale. This is clearly seen in figure \ref{fig:GW_vs_HTL}, where we computed the full HTL and GW potentials with the prescription given by \eqref{eq:GW_HTL_match}. It is clear from this numerical exercise, that up to around $|\x|\sim \frac{2}{m_{\rm D}}$, the two potentials match almost exactly. As one goes to larger dipoles sizes, the leading logarithmic approximation is not enough and a new map taking into account higher order terms would have to be constructed\footnote{This is however not important, because due to color transparency, such dipole sizes do not give an important contribution to the emission spectrum \eqref{eq:spectrum_general}.}. Therefore, for the rest of this paper we will work with the the leading order accuracy cross section given by
\begin{equation}\label{eq:scattering_cross_section}
\frac{\rmd^2\sigma}{\rmd^2\k}=\frac{g^4n(t)}{\k^4}\, .
\end{equation}
The propagator potential (at leading logarithmic accuracy) then reads
\begin{equation}\label{eq:v}
v(\x,t)=C_A\int_\k     \frac{d^2\sigma}{d^2\k} (1-e^{i\k\cdot\x})\equiv \frac{1}{4}\hat{q}(\x^2,t)\x^2=\frac{1}{4}\hat{q_0}\x^2 \log\left(\frac{1}{\mu^{\star2}\x^2}\right) \, ,
\end{equation}
where $\hat{q}$ is the medium transport coefficient and 
$\mu^\star$ is introduced as an infrared cut-off. Note that in general $\hat{q}$ has a trivial dependence on time via $n$, which in this paper we take to be $n(t)=n\Theta(L-t)$, with $L$ the medium length (plasma brick model). \par

\subsection{The harmonic oscillator approximation}
In general the solution to equation \eqref{eq:cK_Sch} is not known in a closed form even for the leading logarithmic potential \eqn{eq:v}, but the case of vacuum propagation and as we shall see shortly, the harmonic oscillator, admit a complete full analytic treatment. The free propagator reads
\begin{equation}\label{eq:K_0}
\cK_0(\x,t_2|\y,t_1)=\frac{\omega}{2\pi i (t_2-t_1)}\exp\left[\frac{i\omega (\x-\y)^2}{2(t_2-t_1)}\right]\,.
\end{equation}
To connect \eqn{eq:v} with the harmonic oscillator, following the BDMPS-Z approach, one assumes that $1/\x^2\sim Q^2$, where $Q^2$ is the typical transverse scale of the process to be determined later. This makes the potential that of a harmonic oscillator (HO). Therefore a  closed form solution  for $\cK\equiv\cK_{\rm HO}$ exists \cite{Stegun}
\begin{equation}\label{eq:cK_BDMPS}
K_{\rm HO}(\x,t|\y,t_1)= \frac{\omega}{2\pi i S(t,t_1)}\exp\left[\frac{i\omega}{2S(t,t_1)} \left\{ C(t_1,t)\,\x^2+C(t,t_1)\,\y^2-2\x\cdot\y\right\}\right]\, .
\end{equation}
The functions $S$ and $C$ are the solutions to the initial condition problems
\begin{equation}\label{eq:Abel}
\begin{split}
\left[\frac{d^2}{d^2t}+\Omega[t]\right]S(t,t_0)&=0 \, ,\quad S(t_0,t_0)=0 \,,\quad \partial_t  S(t,t_0)_{t=t_0}=1 \, , \\   
\left[\frac{d^2}{d^2t}+\Omega[t]\right]C(t,t_0)&=0 \, ,\quad C(t_0,t_0)=1 \,,\quad \partial_t C(t,t_0)_{t=t_0}=0\, .
\end{split}
\end{equation}
Here we have defined the (complex) frequency of the harmonic oscillator 
\begin{equation}\label{eq:BigOmega}
\Omega(t)=\frac{1-i}{2}\sqrt{\frac{\hat{q}(t)}{\omega}}\, ,
\end{equation}
and 
\begin{equation}\label{eq:qhat_HO}
\hat{q}(t)=\hat{q}_0(t)\log\left(\frac{Q^2}{\mu^{\star2}}\right)    \, .
\end{equation}
In the case where the medium is a plasma brick we have the closed form solutions for $C$ and $S$ given by 
\begin{equation}
S(t,t_0)=\frac{1}{\Omega}\sin(\Omega(t-t_0)) \quad , \quad C(t,t_0)=\cos(\Omega(t-t_0))\,.
\end{equation}
The time dependence on $\Omega$ has disappeared since the medium is assumed to be static.\par 
In general, the functions $C$ and $S$ can be obtained for any medium, by either solving the path integral $\cK$ via the semi-classical approximation \footnote{In fact, for quadratic Lagrangians this approximation gives the exact result \cite{Stegun}.} (Pauli's formula) \cite{Apolinario:BDMPS_full} or by following the Wronskian approach (i.e. solving \eqref{eq:Abel}; see \cite{Paper1,Arnold_Simpleformula} for details). For the present paper, the second approach is more useful.\par 
The properties of the functions $C$ and $S$ can be studied in general for any medium (see \cite{Arnold_Simpleformula}). In particular, one can show that they obey the following identity \cite{Arnold_Simpleformula,Paper1,Paper2}
\begin{equation}\label{eq:CS_infinity}
\frac{C_{\infty, s}}{S_{\infty, s}}=-\frac{\partial_s C_{s,L}}{C_{s,L}}= \Omega^2(s) \frac{S_{s,L}}{C_{s,L}}    \, ,
\end{equation}
where we also introduced the handy notation $C(t,s)\equiv C_{t,s}$ and equivalently for $S$. In the future we will extend this notation to allow the shorthand writing of $C(t_2,t_1)\equiv C_{2,1}$, with the same applying to $S$ and only valid when the dependency in $t$ is clear and only the sub-indices matter. Although not generally true, we will also treat $C$ has being an even function, which is true in the plasma brick model.

\subsection{The general structure of the spectrum}\label{sub_sec:the_general_structure_of_the_spectrum}
The Improved Opacity Expansion is realized by expanding the full medium induced spectrum around the BDMPS-Z solution, such that the leading order (LO) term in the expansion matches the known solution and the higher order (N$^m$LO) terms take into account the hard scattering contributions. This is achieved by rewriting the potential $v$ as \cite{Paper1}
\begin{equation}\label{eq:v_expansion}
v(\x,t)=\frac{1}{4}\x^2\log\left(\frac{1}{\mu^{\star2}\x^2}\right)=\frac{1}{4}\x^2\left(\log\left(\frac{Q^2}{\mu^{\star2}}\right)+\log\left(\frac{1}{Q^2\x^2}\right)\right)\equiv v_{\rm HO}(\x,t)+\delta v(\x,t)\,  .
\end{equation}
 Here $Q^2$ is the matching scale between the two regimes of the spectrum and as it is clear from \eqref{eq:v_expansion}, the spectrum is independent of it, when all orders in perturbation theory are taken into account. For the moment $Q^2$ is assumed to be arbitrary, but as we shall see the logarithmic structure of the expansion requires a specific choice that is as expected the typical transverse momentum acquired by the radiated gluon $Q^2\sim \sqrt{\omega \hat q }$.   \par
This expansion can be incorporated into $\cK$ by using the Dyson-like equation for the propagator
\begin{equation}\label{eq:K_Dyson}
\cK(\x,t,\y,s)=-\int_\z\int_{s}^t du \ \cK_{\rm HO}(\x,t|\z,u)\delta v(\z,u)\cK(\z,u|\y,s)\, ,
\end{equation}
where $\int_\z \equiv \int d^2\z$. Each order in perturbation theory is then obtained by expanding the above equation in powers of $\hat{q}_0$ (see the detailed discussion in \cite{Paper1}). Doing this procedure and using equation \eqref{eq:spectrum_general} the full spectrum reads\footnote{The truncation of these series leads to a dependency on the matching scale $Q^2$.} \cite{Paper1,Paper2}
\begin{equation}\label{eq:full_spectrum}
\begin{split}
\omega \frac{\rmd I}{\rmd\omega}&=\omega \frac{\rmd I^{ \rm HO=LO}}{\rmd\omega}+ \omega \frac{\rmd I^{\rm NLO}}{\rmd\omega}+\cdots  =\omega  \frac{\rmd I^{\rm LO}}{\rmd\omega}+ \sum_{m=1}^\infty\omega \frac{\rmd I^{{\rm{ N}}^m\rm{ LO}}}{\rmd\omega}\, .
\end{split}
\end{equation}

\subsection{The leading order (BDMPS-Z) term}\label{sub_sec:The_leading_order_term}

The leading order term is the well known BDMPS-Z result that can be obtained by using equations \eqref{eq:cK_BDMPS} and \eqref{eq:K_0} in the general formula for the spectrum \eqref{eq:spectrum_general}. One then obtains the compact formula \cite{Paper1}
\begin{equation}
 \omega \frac{\rmd I^{\rm LO}}{\rmd\omega}= -2\Bar{\alpha}\Re \left[\int_0^\infty dt_2 \int_0^{t_2} dt_1 \frac{1}{S^2(t_2,t_1)}-\frac{1}{(t_2-t_1)^2}\right] \ ,   
\end{equation}
where from this point on we always assume that we are working within the plasma brick model. It is then easy to show that the $S$ and $C$ functions obey the following differential relation \footnote{In fact, as is shown in \cite{Paper1,Arnold_Simpleformula}, this property holds for all medium models and not just for the case of the plasma brick case.}
\begin{equation}
\partial_t\left(\frac{C(t,t_0)}{S(t,t_0)}\right)=-\frac{1}{S^2(t,t_0)}  \, .
\end{equation}
This allows one to make the $t_2$ integration directly and obtain
\begin{equation}
\begin{split}
 \omega \frac{\rmd I^{\rm LO}}{\rmd\omega}&= -2\Bar{\alpha}\Re \left[\int_0^{\infty} dt_1 \ \frac{C(t_1,t_1)}{S(t_1,t_1)}-\frac{C(\infty,t_1)}{S(\infty,t_1)}-\frac{1}{t_1-t_1}\right] 
 \\&=2\Bar{\alpha}\Re \left[\int_0^{\infty} dt_1 \ \frac{C(\infty,t_1)}{S(\infty,t_1)}\right] =2\Bar{\alpha}\Re \left[\int_0^{\infty} dt_1 \ -\frac{\partial_{t_1}C(t_1,L)}{C(t_1,L)}\right]
 \\&=\log C(0,L)\, ,
 \end{split}
\end{equation}
where in the first step we cancelled the divergent pieces between $\cK_{\rm HO}$ and $\cK_0$ and we have used \eqref{eq:CS_infinity} in the last step.\par
Then the spectrum finally reads
\begin{equation}\label{eq:LO_term}
\omega \frac{\rmd I^{\rm LO}}{\rmd\omega}=2\Bar{\alpha}\log|\cos(\Omega L)|   \, , 
\end{equation}
where $\Bar{\alpha}=\alpha_s C_R/\pi$. Defining the characteristic  frequency $\omega_c$ as the typical frequency of the emitted gluon with formation time of the order of the medium length
\begin{equation}\label{eq:wc}
\omega_c=\frac{1}{2}\hat{q}L^2 \, ,  
\end{equation}
we can obtain the asymptotics of \eqref{eq:LO_term}
\begin{equation}\label{eq:LO_term_asymptotic}
\omega \frac{\rmd I^{\rm LO}}{\rmd\omega}=2\Bar{\alpha}
\
    \begin{dcases}
        \sqrt{\frac{\omega_c}{2\omega}} & , \ \omega\ll\omega_c \\
        \frac{1}{12}\left(\frac{\omega_c}{\omega}\right)^2 & , \ \omega \gg \omega_c \\
    \end{dcases}
\     \, ,
\end{equation}
which quantitatively shows the scalings discussed in section \ref{sec:intro}. As we shall see the HO does not correctly capture the scaling when $\omega \gg \omega_c$, i.e., $\omega^{-1}$, which is dominated by a single hard scattering. 

\subsection{The $m^{th}$ order correction }\label{sub_sec:The_M_order_correction}

The general form for the $m^{th}$ contribution to the full spectrum which includes the hard scattering potential is given by
\begin{equation}\label{eq:NMLO}
\begin{split}
\omega    \frac{\rmd I^{{\rm{ N}}^m\rm{LO}}}{\rmd\omega}&=(-1)^m\frac{\Bar{\alpha}\pi}{\omega^2}2\Re\bigg[ \int_{0}^\infty dt_2\int_{0}^{t_2} dt_1 \  \int_{\z_1}\int_{\z_2}\cdots \int_{\z_{ m}} \int_{t_1}^{t_2} ds_{ m} \int^{s_{ m}}_{t_1} ds_{ m-1} \cdots \int_{t_1}^{s_{ 2}} ds_1 
\\& \times \bdel_\x \cdot \bdel_\y \
\cK_{\rm HO}(\x,t_2; \z_{m},s_{ m})\delta v(\z_{ m},s_{ m}) \cK_{ \rm HO}(\z_{ m},s_{ m};\z_{ m-1},s_{ m-1})\delta v(\z_{ m-1},s_{ m-1})
\\&\times \cK_{\rm HO}(\z_{m-1},s_{ m-1};\z_{ m-2},s_{ m-2})\cdots \times  \cK_{\rm HO}(\z_1,s_1; \y,t_1)
\bigg]_{\x=\y=0}    \, .
\end{split}    
\end{equation}
Here we have ordered the times of each scattering center from $t_1$ to $t_2$ in increasing order of the sub-index, running from $1$ to $m$. The transverse position of the $i^{th}$ scattering center $\z_i$ is also ordered from the first scattering center ($\z_1$) to the last one ($\z_{ m}$).\par 
Equation \eqref{eq:NMLO} is obtained by iteratively using equation \eqref{eq:K_Dyson} in equation \eqref{eq:spectrum_general}. As was shown in \cite{Paper1}, the two extreme propagators can be integrated out, after performing the derivatives and using the general relation
\begin{equation}\label{eq:time_integrations}
\begin{split}
&\int_0^\infty dt_2\int_0^{t_2}dt_1 \int_{t_1}^{t_2}ds_{ m}\int_{t_1}^{s_{ m}}ds_{ m-1} \cdots \int_{t_1}^{s_{2}}ds_1=
\\=&\int_0^\infty ds_1 \int_{s_1}^\infty ds_{2} \cdots \int_{s_{m-1}}^\infty ds_{ m}\int_{s_{m}}^\infty dt_2\int_0^{s_1}dt_1  \, .
\end{split}
\end{equation}
We are then left with just the intermediate position integrals and the time integrations at each scattering center. Introducing the representation for $\cK_{\rm HO}$ in \eqref{eq:cK_BDMPS} and using the explicit formula for $\delta v$ one eventually obtains the compact formula
\begin{equation}\label{eq:NMLO_final}
\begin{split}
\omega    \frac{\rmd I^{{\rm{N}}^m\rm{LO}}}{\rmd\omega}&=\frac{\Bar{\alpha}\hat{q}_0^m}{2^{3m-2}\pi^{m}}\Re\bigg[ \left[\frac{\z_1\cdot\z_m}{\z_1^2\z_m^2}\right]\prod_{j=1}^m \fint_{\z_j} \int_{s_{j-1}}^{L}ds_j \ \z_j^2 \log\left(\frac{1}{Q^2\z_j^2}\right)
 \\&\times\sigma_{j+1,j}\exp\left[k_j^2\z_j^2\right]\exp\left[-\sigma_{j+1,j}\z_{j+1}\cdot\z_j\right]\bigg] \, , 
\end{split}    
\end{equation}
where we use the prescriptions: $s_{0}=0$, $\sigma_{ m+1,  m}=1$ and $\z_{ m+1}=0$. Also, the factor depending on $\z_{ m}$ and $\z_1$ outside the product, should be understood to be integrated over (i.e. the factor enters the $\z_1$ and $\z_{m }$ integrals; this is denoted by the slashed integral symbol). Here the factor $\pi^m$ comes from the m factors of $\cK_{\rm HO}$ present in the general formula and the factor $\hat{q}_0^{m}$ is due to the presence of m $\delta v$ terms. The $2^{3 m}$ appears as a combination of the $\cK_{\rm HO}$ normalisation factors and the terms in $\delta v$. \par 
We have introduced the following functions
\begin{equation}\label{eq:kj}
k_j^2=\frac{i\omega}{2}\left[\frac{C_{j,j-1}}{S_{j,j-1}}+\frac{C_{j+1,j}}{S_{j+1,j}}\right]    \, ,
\end{equation}
with the boundary properties $C_{1,0}=C_{\infty,1}$ and $C_{m+1,m}=C_{m,0}$ and the same for the $S$ function. Also
\begin{equation}
\sigma_{k,j}=\frac{i \omega}{S_{k,j}}   \, . 
\end{equation}

It is clear from equation \eqref{eq:NMLO_final}
that performing the remaining integrations is non trivial when $m\geq3$, so that the NLO and NNLO are special cases where one can hope to make analytical simplifications. For completeness we give the NLO ($m=1$) term, already computed in \cite{Paper1,Paper2}
\begin{equation}\label{eq:NLO_term}
\omega    \frac{\rmd I^{\rm NLO}}{\rmd\omega}=\frac{\Bar{\alpha}\hat{q}_0}{2\pi}\Re\bigg[ \int_{\z} \int_{0}^{L}ds \ \log\left(\frac{1}{Q^2\z^2}\right)
\exp\left[k^2(s)\z^2\right]\bigg]   \, ,
\end{equation}
with
\begin{equation}\label{eq:k_NLO}
k^2(s)=\frac{i\omega}{2}\left[\frac{C_{1,0}}{S_{1,0}}+\frac{C_{2,1}}{S_{2,1}}\right]= \frac{i\omega}{2}\left[\frac{C_{\infty,s}}{S_{\infty,s}}+\frac{C_{s,0}}{S_{s,0}}\right] \, ,
\end{equation}
where in the second step we have translated from the notation for general $m$ to the case $m=1$ and we used the boundary properties of the $C$ function. This result perfectly matches the result from the previous papers\footnote{We would like to point out that in equation \eqref{eq:k_NLO} there is an overall extra minus sign when compared to \cite{Paper1,Paper2}. This corrects the small mistake present previously, which does not affect the results significantly.}.

\subsection{The Next-to-Leading order correction}\label{sec:Next_to_Leading_Order_correction}
Before computing the NNLO term in the IOE, we present the NLO contribution already computed in previous work \cite{Paper1}.\par 
Starting from \eqref{eq:NLO_term} we use the identity 
\begin{equation}
\int_0^\infty du \log\left(\frac{1}{u}\right) e^{-bu}=\frac{1}{b}\left(\log(b)+\gamma_E\right)    \, ,
\end{equation}
to get the spectrum
\begin{equation}\label{eq:NLO_local}
\omega \frac{\rmd I^{\rm LO}}{\rmd\omega}=\frac{1}{2}\bar{\alpha}\hat{q}_0 \Re\left[\int_0^L ds\ \frac{-1}{k^2(s)}\left(\log\left(-\frac{k^2(s)}{Q^2}\right)+\gamma_E\right)\right]   \, ,
\end{equation}
where the angular integration was also carried out. \par 
In analogy to what was done for the LO term, we also study the limiting cases $\omega\to0$ and $\omega\to\infty$.\par In the first case, it is easy to check that $k^2(s)\to -\omega \Omega$. The NLO contribution can then be computed exactly \cite{Paper1} and reads
\begin{equation}\label{eq:NLO_smallw_maineq}
\begin{split}
\lim_{\omega \to 0} \omega \frac{\rmd I^{\rm NLO}}{\rmd\omega} &=\frac{\Bar{\alpha}}{2}\hat{q}_0\Re\bigg[\int_0^L \frac{2}{(1-i)\sqrt{\omega \hat{q}}}\left(\log\left(\frac{(1-i)\sqrt{\omega \hat{q}}}{2Q^2}\right)+\gamma_E\right)\bigg]
\\&= \frac{\Bar{\alpha}}{2}\left(\frac{\hat{q}_0}{\hat{q}}\right)\sqrt{\frac{\hat{q}L^2}{\omega}}\left[\gamma_E+\log\left(\frac{\sqrt{\omega \hat{q}}}{\sqrt{2}Q^2}\right)+\frac{\pi}{4}\right]\sim \omega \frac{dI^{\rm LO}}{d\omega}\left(\frac{\hat{q}_0}{\hat{q}}\right) \, ,
\end{split}
\end{equation}
which shows that at the low frequency part of the spectrum this contribution scales like the LO term but suppressed by a logarithmic contribution $\sim\log^{-1}\left(\frac{\sqrt{\hat{q}\omega}}{\mu^2}\right)$. To get to this result we have assumed in the last step that $Q^2\sim\sqrt{\hat{q}\omega}$.\par 
On the other hand, the high energy limit implies that $k^2(s)\to \frac{i\omega}{2s}$.
The NLO term becomes dominant in this region of phase space and we have from equation \eqref{eq:NLO_local}
\begin{equation}\label{eq:NLO_scaling_high_energy}
\lim_{\omega \to \infty}\omega \frac{\rmd I^{\rm NLO}}{\rmd\omega}\sim \Bar{\alpha}\hat{q}_0\frac{\pi}{4}\frac{L^2}{2\omega}=\frac{\Bar{\alpha} \hat{q}_0L}{\mu^{\star2}}\frac{\pi}{4}\frac{\bar{\omega}_c}{\omega}=\frac{\pi}{4}\chi\, \Bar{\alpha} \frac{\bar{\omega}_c}{\omega} \, ,
\end{equation}
which matches the asymptotic behavior of GLV \cite{Paper1,GLV}. Here we used $\bar{\omega}_c\equiv\frac{\mu^{\star2}L}{2}$ and $\chi \equiv\frac{\hat{q}_0L}{\mu^{\star2}}$ that measures the opacity. This term is dominant compared to LO contribution (the BDMPS-Z result is power suppressed).\par

\section{The Next-to-Next-to-Leading order correction}\label{sec:The_Next_to_Next_to_Leading_order}

Using \eqref{eq:NMLO_final} we can obtain the NNLO term directly. The angular integrations can be done in a straightforward way and in the end one is left with 4 integrations to perform.
\begin{equation}\label{eq:NNLO_final}
\begin{split}
\omega \frac{\rmd I^{\rm NNLO}}{\rmd\omega}&=-\frac{\Bar{\alpha}}{4}\Re\bigg[\hat{q}_0^2\int_0^L ds_2 \int_{s_2}^L ds_1   \ \sigma_{s_1,s_2}\int_{z_1z_2} \log\left(\frac{1}{Q^2z_1^2}\right)\log\left(\frac{1}{Q^2z_2^{2}}\right)z_1^2z_2^{2}
\\&\times e^{k_1^2z_1^2}e^{k_2^2z_2^{2}}J_1(z_1 z_2 \sigma_{s_1,s_2})\bigg] \, ,
\end{split}    
\end{equation}
where $J_1$ is the Bessel of the first kind of degree $1$ and $\int_z\equiv\int_0^\infty dz$. From this point on the indices in $\sigma$ will be dropped. We have
\begin{equation}
k_1^2=\frac{i\omega}{2}\left(\frac{C_{1,2}}{S_{1,2}}+\frac{C_{\infty, 1}}{S_{\infty, 1}}\right) \quad , \quad k_2^2=\frac{i\omega}{2}\left(\frac{C_{1,2}}{S_{1,2}}+\frac{C_{2,0}}{S_{2,0}}\right) \quad , \quad \sigma=\frac{i\omega}{S_{1,2}} \, .
\end{equation}
Formally, it is still possible to further simplify \eqref{eq:NNLO_final} by performing one of the $z$ integrations. This leads to the appearance of a finite sum of Bessel functions and a logarithmic contribution. Although this decreases the number of integrations by one, the result obtained is neither suitable for numerical implementation nor is it of easy analytic study.\par 
We proceed to analyze equation \eqref{eq:NNLO_final} in two limiting regimes. First we explore the region where $\omega\to \infty$, i.e. where the major contribution to the spectrum should come from the NLO term. Then we study the opposite limit where $\omega \to 0$.

\subsection{Large frequency limit}\label{sub-sec:Large_frequency_limit}
In this regime we let $\omega \to \infty \equiv \Omega \to (1-i)\times 0$. We notice that in such regime the $k_1$, $k_2$ and $\sigma$ functions can be simplified using the fact that $C_{s_a,s_b}\to 1$ and $S_{s_a,s_b}\to s_a-s_b$
\begin{equation}\label{eq:k1_k2_sigma_high_energy}
\sigma \to \frac{i \omega}{s_1-s_2} \quad,\quad k_1^2\to \frac{i \omega}{2(s_1-s_2)} \quad,\quad k_2^2\to \frac{i \omega}{2}\frac{s_1}{s_2(s_1-s_2)} \, .
\end{equation}
Before using this approximation, we rewrite the spectrum in such a way that the $z_1$, $z_2$ can be integrated out. We proceed by power expanding the Bessel function 
\begin{equation}\label{eq:NNLO_J1_expansion}
\begin{split}
\omega \frac{\rmd I^{\rm NNLO}}{\rmd\omega}&=-\frac{\Bar{\alpha}}{8}\hat{q}_0^2\Re\bigg[\int_0^L ds_2 \int_{s_2}^L ds_1   \ \sum_{n=0}^\infty\frac{(-1)^n}{n!(n+1)!} \sigma^{2(n+1)} \left(\frac{1}{4}\right)^n 
\\&\times \int_{z_1z_2}\log\left(\frac{1}{Q^2z_1^2}\right)\log\left(\frac{1}{Q^2z_2^{2}}\right) z_1^{2n+3}z_2^{2n+3} e^{k_1^2z_1^2}e^{k_2^2z_2^{2}}\bigg] \, .
\end{split}    
\end{equation}
This representation is advantageous since it allows to directly do the integrations in $z_1$ and $z_2$ by using 
\begin{equation}\label{eq:integration_trick}
\int_x \log\left(\frac{1}{Q^2x^2}\right)z^{2n+3}e^{k^2x^2}=\frac{(n+1)!}{2}\left(-\frac{1}{k^2}\right)^{n+2} \log\left(-\frac{k^2}{Q^2 E\psi(n+2)}\right) \, ,
\end{equation}
where $E\psi(n)=\exp(\psi(n))$, $\psi(n)=\Gamma' (n)/\Gamma(n)$ and $\Gamma$ is the gamma function. Putting all together the spectrum then reads
\begin{equation}\label{eq:NNLO_large_w}
\begin{split}
\omega \frac{\rmd I^{\rm NNLO}}{\rmd\omega}&=\frac{\omega}{32}\hat{q}_0^2 \Im \bigg[\int_0^L ds_2 \int_{s_2}^L ds_1   \ \frac{\sigma}{S_{12}}\left(\frac{1}{k_1^2k_2^2}\right)^2\sum_{n=0}^\infty\frac{(-1)^n(n+1)}{4^n} \sigma^{2n}  
\\&\times  \left(\frac{1}{k_1^2k_2^2}\right)^n\log\left(-\frac{k_1^2}{Q^2E\psi(n+2)}\right)\log\left(-\frac{k_2^2}{Q^2E\psi(n+2)}\right)
\bigg] \, .
\end{split}  
\end{equation}
This achieves our goal of removing the integrations in the intermediate positions. However, we are left with an infinite series, which might not converge for all the parameter space. We notice that formally, the sequence being summed scales with 
\begin{equation}\label{eq:scaling_high_energy}
\sim\frac{n+1}{4^n}\left(\frac{\sigma^2}{k_1^2k_2^2}\right)^n \psi(n+2)\psi(n+2)\stackrel{n\gg1}{\sim} n\left(\frac{\sigma^2}{4k_1^2k_2^2}\right)^n \log^2 (n)  \, .
\end{equation}
We see that the converge of the series is then controlled by the dimensionless quantity $\frac{\sigma^2}{4k_1^2k_2^2}$. We notice that in the high energy limit using \eqref{eq:k1_k2_sigma_high_energy}, equation \eqref{eq:scaling_high_energy} reduces to 
\begin{equation}\label{eq:scaling_high_energy}
n\left(\frac{\sigma^2}{4k_1^2k_2^2}\right)^n \log^2 (n) \sim \left(\frac{s_2}{s_1}\right)^n n \log^2 (n)  \, ,
\end{equation}
which is a well posed expansion parameter since $s_2<s_1$. Therefore, in the high energy limit, the power expansion representation of the spectrum gives a convergent series, and therefore only a finite number of terms in the expansion are required to achieve a reasonable numerical convergence. \par 
Before numerically solving equation \eqref{eq:NNLO_large_w}, let us study the asymptotic behavior of the spectrum analytically. The LO term for $\omega\gg \omega_c$ scales with $\Bar{\alpha}(\frac{\omega_c}{\omega})^2$ (see \eqref{eq:LO_term_asymptotic}). The NLO term is given by \eqref{eq:NLO_scaling_high_energy} and scales with $\frac{\Bar{\omega}_c}{\omega}$ and is therefore the leading order term.\par  
The NNLO term is better discussed in terms of the rate $\omega \frac{\rmd I^{\rm NNLO}}{\rmd L\rmd\omega}$. Then in the same limit as before the contribution reads 
\begin{equation}\label{eq:NNLO_rate_large_omega}
\begin{split}
\lim_{\omega \to \infty}\omega \frac{\rmd I^{\rm NNLO}}{\rmd L\rmd\omega}&=\frac{\Bar{\alpha}}{2\omega^2}\hat{q}_0^2 \Re \bigg[\int_0^L ds_2   \ \left(\frac{s_2}{L}\right)^2 (L-s_2)^2
 \sum_{n=0}^\infty (-1)^n(n+1) \left(\frac{s_2}{L}\right)^n \\&\times\log\left(-i\frac{\omega}{2(L-s_2)Q^2 E\psi(n+2)}\right)\log\left(-i\frac{\omega L}{2s_2(L-s_2)Q^2 E\psi(n+2)}\right)
\bigg]  \, .
\end{split}  
\end{equation}
Taking the real part we have
\begin{equation}
\begin{split}
\lim_{\omega \to \infty}\omega \frac{\rmd I^{\rm NNLO}}{\rmd L\rmd\omega}&=\frac{\Bar{\alpha}}{2\omega^2}\hat{q}_0^2  \bigg[\int_0^L ds_2 \ \left(\frac{s_2}{L}\right)^2 (L-s_2)^2
 \sum_{n=0}^\infty (-1)^n(n+1) \left(\frac{s_2}{L}\right)^n \\&\times\left(\log\left(\frac{\omega}{2(L-s_2)Q^2 E\psi(n+2)}\right)\log\left(\frac{\omega L}{2s_2(L-s_2)Q^2 E\psi(n+2)}\right)-\frac{\pi^2}{4}\right)
\bigg]  \, .
\end{split}  
\end{equation}
To proceed we rescale the time integration with $u=s_2/L$ and only keep the leading order contribution in the logarithms $\sim \log(\frac{\omega}{Q^2L})$.
\begin{equation}\label{eq:final_NNLO_large_omega_scaling}
\begin{split}
\lim_{\omega \to \infty}\omega \frac{\rmd I^{\rm NNLO}}{\rmd L\rmd\omega}&= \frac{\Bar{\alpha}L^3}{2\omega^2}\hat{q}_0^2 \int_0^1 du \ u^2(1-u)^2
 \sum_{n=0}^\infty (-1)^n(n+1) u^n \\&\times\log\left(\frac{\omega}{2L(1-u)Q^2}\right)\log\left(\frac{\omega }{2Lu(1-u)Q^2}\right)
 \\&\sim\frac{\Bar{\alpha}}{L}\chi \left(\frac{\bar{\omega}_c}{\omega}\right)^2\log^2\left(\frac{\omega}{Q^2L}\right)  \, ,
\end{split}  
\end{equation}
where we have neglected all terms not doubly enhanced by logarithms and the remaining (finite) numerical factor coming from the integration in $u$. In the last step the $u$ dependence in the logarithms can be dropped since it is only single logarithmic enhanced. \par 
The full spectrum predicted by the IOE is then dominated by the NLO term to all orders, since all higher order terms contribute with power corrections $\sim \frac{\Bar{\omega}_c}{\omega}$, which are suppressed. There are also some logarithmic enhancements, but this are always small compared to the power terms. Notice that, when moving away from the strict high energy limit, the NLO (i.e. leading term) will originate corrections (coming from the $k^2$ expansion) which contribute at LO order $\sim \left(\frac{\omega_c}{\omega}\right)^2$, with some possible logarithmic corrections. This also applies to higher contributions, where the corrections coming from the high energy limit expansion of the $k$'s and $\sigma$'s functions come with extra power law contributions. This fact ensures that the NLO term will always be the dominant piece in the IOE. \par 
\begin{figure}%
    \centering
    {{\includegraphics[width=.45\linewidth]{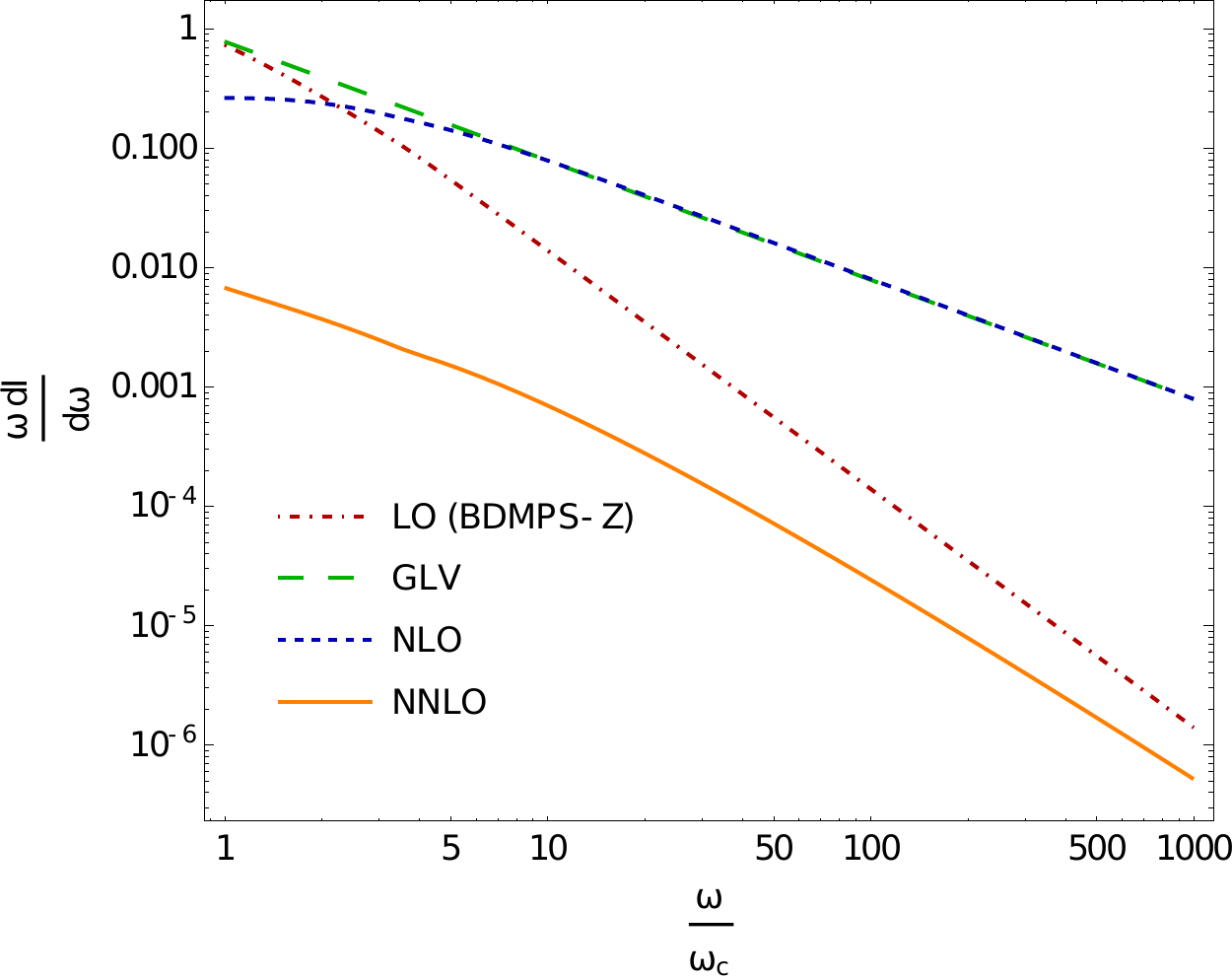}
    }}%
    \qquad
   {{\includegraphics[width=.45\linewidth]{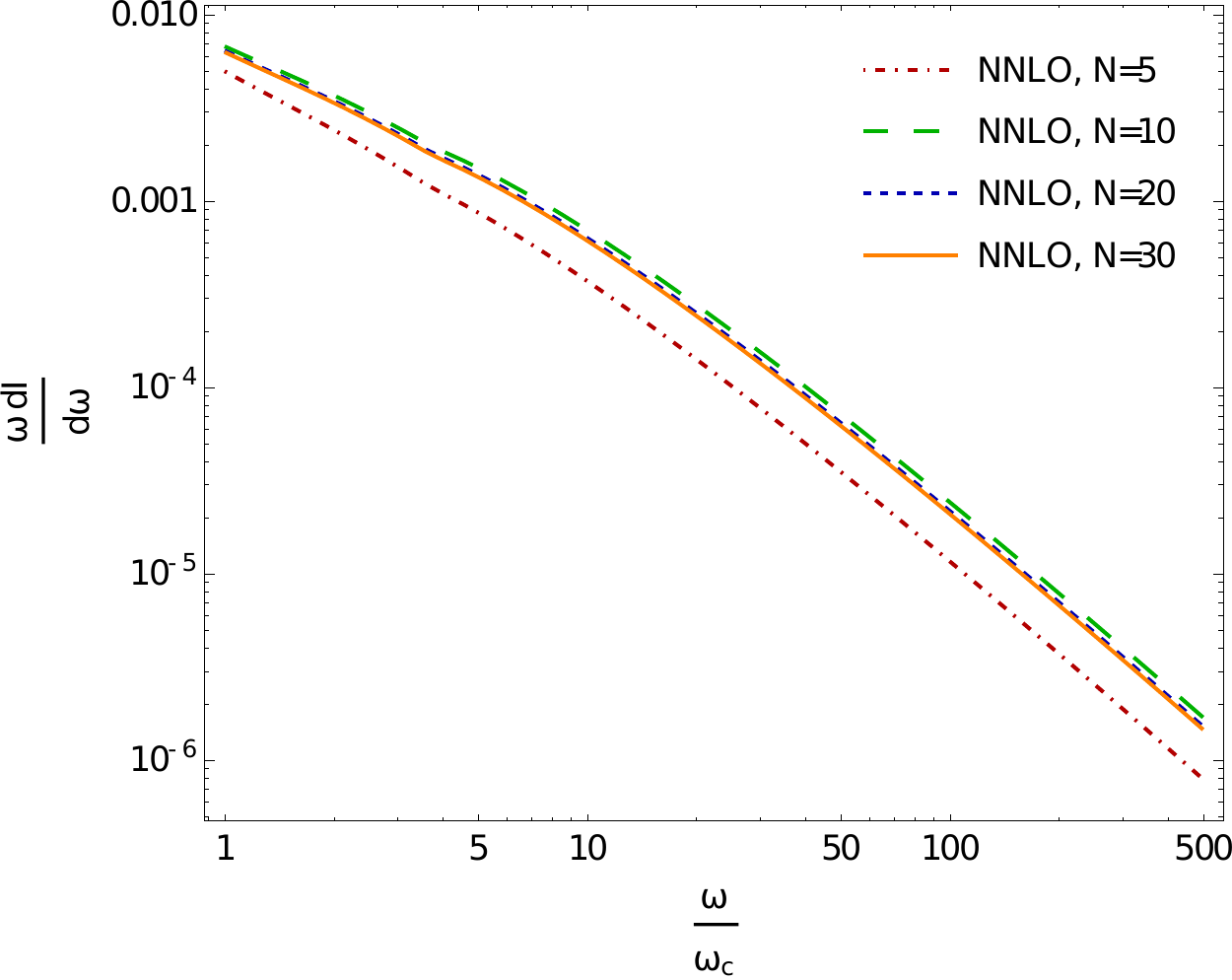}
    }}%
    \caption{\textbf{Left:} The different contributions to the improved opacity expansion spectrum (LO, NLO and NNLO) and the GLV spectrum, in the high frequency regime ($\omega\gtrsim \omega_c$). The plotted curves are given with the overall constant $\Bar{\alpha}=1$. We use the following set of numerical parameters: $\hat{q}_0=0.1$ GeV$^3$, $\mu^\star=0.2$ GeV and $L=6 $ fm. This set of parameters is used for the rest of the numerical results, unless otherwise stated. \textbf{Right:} The NNLO term computed using equation \eqref{eq:NNLO_large_w}, while replacing the upper limit of the sum by N$=5$, N$=10$, N$=20$ and N=$30$. The plots that follow in the rest of this paper use N$=10$, since it shows an extremely good convergence and small computational time.}%
    \label{fig:LO_GLV_NLO_NNLO}%
\end{figure}

In figure \ref{fig:LO_GLV_NLO_NNLO} (\textbf{Left}) we present the numerical computation of the LO, NLO (already shown in \cite{Paper1}) and the NNLO terms in the IOE. In addition, we present the GLV spectrum. The NNLO term is obtained by direct numerical implementation of equation \eqref{eq:NNLO_large_w}, and thus this result is only valid for sufficiently large $\omega$ (in this case, we summed the first $11$ terms of the series; see figure \ref{fig:LO_GLV_NLO_NNLO} (\textbf{Right}) for the comparison of different truncation values.).  \par 
The numerical results depict exactly what was argued before. At large $\omega$, the NLO term becomes the dominant contribution to the spectrum. The NNLO at LO lines become almost parallel at large $\omega$, thus showing that these two terms give the same asymptotic contribution (this is not strictly true, since they will differ by subleading logarithmic terms).  Moreover, we also notice that the actual numerical values assumed by the NNLO curve are at their best only an order of magnitude smaller than the NLO contribution. This shows, that for practical purposes, in this regime, the NLO truncation already offers an excellent approximation to the full spectrum, and subleading corrections do not change the behavior of the IOE.\par

\subsection{Small frequency limit}\label{sub-sec:Small_frequency_limit}
The small frequency regime requires a more delicate approach. This is mainly due to the fact that in this limit the BDMPS-Z solution, without any kinematic constraints \cite{Quenching_weights,Jet_tomography}, is divergent. In the case of the LO and NLO, this divergence is well under control, since the diverging pieces factorize from the remaining terms. This is no longer true at NNLO order, and thus requires a more careful treatment.\par 
Our starting point is again equation \eqref{eq:NNLO_final}, but we now take the limit $\omega\to 0\equiv \Omega\to (1-i)\times \infty$. From the discussion present in the last subsection, it is clear that in this case using the power expansion of $J_1$ directly is not an optimal strategy, since at some point we would be required to resum all terms in these expansion. Therefore, we keep the integrations in $z_1$ and $z_2$, and take the limiting forms for the $C$ and $S$ functions 
\begin{equation}
\lim_{\omega \to 0} \Omega \frac{\cos(\Omega x)}{\sin(\Omega x)}= i \Omega    \quad , \quad
\lim_{\omega \to 0} \frac{1}{\Omega} \frac{\sin(\Omega x)}{\cos(\Omega x)}= -\frac{i}{ \Omega}     \, .
\end{equation}
We apply this approximation in all the $C$ and $S$ terms but the ones that explicitly dependent on the time difference $s_1-s_2$. In such terms, we cannot use the above approximation\footnote{In case this was done, the result obtained would be divergent.} since $\Omega \sim \frac{1}{t_f}$, where $t_f$ is the typical formation of a BDMPS-Z gluon. Since, parametically, the support of the functions depending on the time difference $s_1-s_2$ is of order $t_f$, these type of dependencies have to be kept in full. If neglected, we would be ignoring the part of the support of the function where it is not damped or highly oscillatory. Then one gets 
\begin{equation}
\lim_{\omega\to 0}k_1^2=k_2^2=\frac{i\omega \Omega}{2}\left(i+\frac{C_{12}}{\Omega S_{12}}\right)  \, ,
\end{equation}
and $\sigma$ cannot be simplified. The NNLO contribution to the IOE spectrum reads
\begin{equation}
\begin{split}
\lim_{\omega \to 0}\frac{\rmd I^{\rm NNLO}}{\rmd\omega}&= -\hat{q}_0^2\frac{\Bar{\alpha}}{4}\bigg[ \int_{0}^L ds_1 \int_{0}^{s_1} ds_2 \ \sigma(\Omega(s_1-s_2)) \int_{zz^\prime} \log\left(\frac{1}{Q^2z_1^2}\right)\log\left(\frac{1}{Q^2z_2^2}\right)
\\&\times z_1^2z_2^{2 } \exp\left[\frac{i\omega \Omega}{2}\left(i+\frac{C_{12}}{\Omega S_{12}}\right)(z_1^2+z_2^{2})\right]J_1(z_1z_2 \sigma)
\bigg]  \, .
\end{split}    
\end{equation}
To proceed, we do the change of variables $(s_1, s_2) \to (s_1, \tau=s_1-s_2)$. To continue, we notice, as argued before, that the main contribution to the integral comes from the region $\tau\sim t_f$, and therefore, the dependence of the result on the upper bound of the integral is small. Therefore, we promote the upper bound $L-s_1\to \infty$. The integration in $s_1$ is then trivial and we are left with just one time integration. This approximation is similar to approaches where the medium induced gluon emission is taken in the Markovian (classical) limit, where the all the shower is dominated by decoherent emissions \cite{BlaizotIancuDominguezYacine}. \par 
In this regime, we can rescale $\tau\to t=\sqrt{\frac{\hat{q}}{4\omega}}\tau$, so that the integration is done in terms of dimensionless quantities. Finally, the functions $C$ and $S$ still present in $k_1$, $k_2$ and $\sigma$ have the complex argument $(1-i)t$. Therefore, we Wick rotate the time integration with the transformation $-iT=(1-i)t$. Then the result reads
\begin{equation}
\begin{split}
 \lim_{\omega \to 0} \omega \frac{\rmd I^{\rm NNLO}}{\rmd\omega}&= -2\Bar{\alpha}\frac{\hat{q}_0^2}{\hat{q}^2}\sqrt{\frac{\hat{q}}{\omega}} L\Re \bigg[ \int_{TUV} \ \frac{i}{\sinh(T)}\log\left(\frac{\sqrt{\hat{q}\omega}}{2Q^2V^2}\right)\log\left(\frac{\sqrt{\hat{q}\omega}}{2Q^2U^2}\right)  
\\&\times U^2V^2 J_1\left(UV(1+i)i \frac{1}{\sinh(T)}\right) e^{\frac{-1+i}{2}\left(\coth(T)+1\right)(U^2+V^{ 2})}
\bigg]   \, ,
\end{split}    
\end{equation}
where we also rescale the position integrations with $U=\left(\frac{\hat {q}\omega}{4}\right)^{1/4}z_1$ and $V=\left(\frac{\hat {q}\omega}{4}\right)^{1/4}z_2$. \par 
To make the integral completely dimensionless, we take the scale $Q^2\sim \sqrt{\hat{q}\omega}$ as in \cite{Paper1}. Then the remaining integral can be computed exactly 
\begin{equation}\label{eq:w0_number}
\begin{split}
  &\Re \bigg[ \int_{TUV} \ \frac{-2i}{\sinh(T)}\log\left(\frac{1}{2V^2}\right)\log\left(\frac{1}{2U^2}\right)  
 U^2V^2
 \\&\times J_1\left(UV(1+i)i \frac{1}{\sinh(T)}\right) e^{\frac{-1+i}{2}\left(\coth(T)+1\right)(U^2+V^{ 2})}
\bigg]  \approx 0.0293246  \, .
\end{split}    
\end{equation}
Thus the scaling for the NNLO term at small frequencies reads
\begin{equation}
\begin{split}
 \lim_{\omega \to 0} \omega \frac{\rmd I^{\rm NNLO}}{\rmd\omega} \sim \Bar{\alpha}\left(\frac{\hat{q}_0}{\hat{q}}\right)^2\sqrt{\frac{\hat{q}L^2}{\omega}}= \omega \frac{\rmd I^{\rm LO}}{\rmd\omega}\left(\frac{\hat{q}_0}{\hat{q}}\right)^2  \, .
\end{split}    
\end{equation}
The NLO contribution is given by equation \eqref{eq:NLO_smallw_maineq} and exhibits the same scaling when $Q^2=\sqrt{\omega\hat{q}}$. Unlike the high energy limit, where we observed that moving away from the strict $\omega \to \infty$ limit originated terms which have to be incorporated in the all orders expansion, at small frequencies (and evaluating  $Q^2=\sqrt{\omega\hat{q}}\equiv Q_c^2$) an all orders expansion can be written and reads
\begin{equation}\label{eq:expansion_IOE_small_frequency_Q2_chosen}
\begin{split}
\lim_{\omega \to 0} \omega \frac{\rmd I}{\rmd\omega} &=\omega \frac{\rmd I^{\rm LO}}{\rmd\omega}\left(1+\frac{c_{1,0}}{\log\left(\frac{Q_c^2}{\mu^{\star2}}\right)}+\frac{
c_{2,0}}{\log^2\left(\frac{Q_c^2}{\mu^{\star2}}\right)}+\cdots\right) 
\\&=\omega \frac{\rmd I^{\rm LO}}{\rmd\omega}\left(1+\frac{0.508}{\log\left(\frac{Q_c^2}{\mu^{\star2}}\right)}+\frac{
0.029}{\log^2\left(\frac{Q_c^2}{\mu^{\star2}}\right)}+\cdots\right)  \, ,
\end{split}
\end{equation}
where the spectrum at LO is understood to be taken in the small frequency regime and the coefficients $c_{0,0}\equiv1$, $c_{1,0} $, $c_{2,0}$, $\cdots$, are pure real numbers, computable order by order. Notice that this expression is consistent at all orders, since every term exhibits the same scaling, up to logarithmic enhancements. The sub-indices of the $c$ coefficients comprise two numbers, the first indicating the power of $\log\left(\frac{Q_c^2}{\mu^{\star2}}\right)$ in the expansion. The role of the second index will shortly become evident. \par
The main difference between \eqref{eq:expansion_IOE_small_frequency_Q2_chosen} and the high energy scaling is that in this case the subleading terms in the expansion (in which the LO contribution is the leading term) can become large, due to logarithmic enhancements. In the high energy limit, this was not possible since each sub leading contribution was power suppressed and the logarithmic enhancements were not dominant.\par 
For a general choice of scale $Q^2$  equation \eqref{eq:expansion_IOE_small_frequency_Q2_chosen} can be written as
\begin{equation}\label{eq:expansion_IOE_small_frequency}
\begin{split}
 \lim_{\omega \to 0} \omega \frac{\rmd I}{\rmd\omega} &=\omega \frac{\rmd I^{\rm LO}}{\rmd\omega}\Bigg(1+\frac{c_{1,0}+c_{1,1}\log\left(\frac{\sqrt{\omega \hat{q}}}{Q^2}\right)}{\log\left(\frac{Q^2}{\mu^{\star2}}\right)}
 \\&+\frac{
c_{2,0}-c_{2,1}\log\left(\frac{\sqrt{\omega \hat{q}}}{Q^2}\right)-c_{2,2}\log^2\left(\frac{\sqrt{\omega \hat{q}}}{Q^2}\right)}{\log^2\left(\frac{Q^2}{\mu^{\star2}}\right)}+\cdots\Bigg)  
\\&=\omega \frac{\rmd I^{\rm LO}}{\rmd\omega}\Bigg(1+\frac{0.508+0.5\log\left(\frac{\sqrt{\omega \hat{q}}}{Q^2}\right)}{\log\left(\frac{Q^2}{\mu^{\star2}}\right)}
 \\&+\frac{
0.029-0.026\log\left(\frac{\sqrt{\omega \hat{q}}}{Q^2}\right)-0.028\log^2\left(\frac{\sqrt{\omega \hat{q}}}{Q^2}\right)}{\log^2\left(\frac{Q^2}{\mu^{\star2}}\right)}+\cdots\Bigg) 
\\&\equiv \omega \frac{\rmd I^{\rm LO}}{\rmd\omega} \sqrt{\frac{W( \sqrt{\omega \hat q }/\mu^{\star 2} )}{\log\left(\frac{Q^2}{\mu^{\star2}}\right)}} = \abar \sqrt{ \frac{\hat q_0W( \sqrt{\omega \hat q }/\mu^{\star 2} )}{\omega} }\, ,
\end{split}
\end{equation}
where the LO spectrum is taken at a general scale $Q^2$. In the last line, we have introduced the function $W$ which formally resums the all-order terms. Notice, that although order by order $W$ exhibits a dependence on $Q^2$ the all order result is independent of the choice of the matching scale. We will discuss the properties of $W$ later on. \par
We see that the second index in the $c_{i,j}$ coefficients denotes the expansion in powers of $\log^j\left(\frac{\sqrt{\omega \hat{q}}}{Q^2}\right)$, opposed to the first index which denotes the terms proportional to powers of $\log^{-i}\left(\frac{Q^2}{\mu^{\star2}}\right)$. We have computed the coefficient $c_{1,1}=\frac{1}{2}$ explicitly in section \ref{sec:Next_to_Leading_Order_correction}.\par 
It is interesting to note the role that the matching scale plays in \eqref{eq:expansion_IOE_small_frequency}. First, suppose that we fix the matching scaling at some constant value $Q^2\equiv\hat{q}_0L$, which is a higher momentum scale at $\omega\sim \omega_c$. Then the logarithms scaling with $\frac{Q^2}{\mu^{\star2}}$ are fixed and the evolution with $\omega$ is encoded in the logarithms of $\frac{\sqrt{\omega\hat{q}}}{Q^2}$. This implies, that at small $\omega$, there is a breakdown of the series since while the LO contribution diverges with $\sim \sqrt{\omega_c/\omega}$ the N$^m$LO contributions diverge (the most diverging piece) $\sim \log^m\left(\frac{\sqrt{\omega\hat{q}}}{Q^2}\right)$, where we have neglected (for this discussion) the different power of logarithms in the denominators since they are constant. In fact, we expect that when $\omega \sim \hat{q}_0 L^2 / \log^2\left(\frac{Q^2}{\mu^{\star2}}\right)$ the expansion breaks down. Note that this scale is parametrically much larger than $\omega_{\rm BH} \sim \hat q \ell^2_\mfp$, and thus while the LO term gives a constant contribution all other orders strongly diverge. This clearly shows that the matching scale has to be chosen such that there is a correct interpolation between the GLV and BDMPS-Z limit, which implies that $Q^2\equiv Q_c^2(\omega)\sim \sqrt{\omega \hat q}$. This choice will allow for mutual cancellations between the different orders in the IOE so that the spectrum does not depend on the matching scale when all orders are resumed and the correct spectrum is recovered (while still away from the Bethe-Heitler limit). \par 
From  \eqref{eq:expansion_IOE_small_frequency}, it seems that the N$^{ m+1}$LO contributions can impact the terms at order N$^{ m}$LO. However, let us suppose that we choose a scale $Q^2\equiv a^2Q_c^2$, where $a$ is dimensionless factor that rescales $Q_c^2\sim \sqrt{\omega \hat q}$. Then, to leading logarithmic accuracy, \eqref{eq:expansion_IOE_small_frequency} becomes
\begin{equation}\label{eq:expansion_around_QS}
\begin{split}
  \lim_{\omega \to 0} \omega \frac{\rmd I}{\rmd\omega}&=  \Bar{\alpha} \sqrt{\frac{\hat{q}_0L^2\log\left(\frac{Q_c^2 a^2}{\mu^{\star2}}\right)}{\omega}}\Bigg[1+\frac{1}{2}\frac{c_{1,0}-\log\left(a^2\right)}{\log\left(\frac{Q_c^2 a^2}{\mu^{\star2}}\right)}+\mathcal{O}\left(\log^{-2}\left(\frac{Q^2}{\mu^{\star2}}\right)\right)\Bigg] 
  \\&=\Bar{\alpha} \sqrt{\frac{\hat{q}L^2}{\omega}}\left(1+\frac{1}{2}\frac{\log\left(a^2\right)}{\log\left(\frac{Q_c^2}{\mu^{\star2}}\right)}\right)\Bigg[1+\frac{1}{2}\frac{c_{1,0}-\log\left(a^2\right)}{\log\left(\frac{Q_c^2 }{\mu^{\star2}}\right)}\left(1-\frac{\log\left(a^2\right)}{\log\left(\frac{Q_c^2}{\mu^{\star2}}\right)}\right)
  \\&+\mathcal{O}\left(\log^{-2}\left(\frac{Q^2}{\mu^{\star2}}\right)\right)\Bigg] 
  \\&=\Bar{\alpha} \sqrt{\frac{\hat{q}L^2}{\omega}}\Bigg[1+\frac{1}{2}\frac{c_{1,0}-\log\left(a^2\right)+\log\left(a^2\right)}{\log\left(\frac{Q_c^2 }{\mu^{\star2}}\right)}
  +\mathcal{O}\left(\log^{-2}\left(\frac{Q^2}{\mu^{\star2}}\right)\right)\Bigg] 
  \\&=\Bar{\alpha} \sqrt{\frac{\hat{q}L^2}{\omega}}\Bigg[1+\frac{1}{2}\frac{c_{1,0}}{\log\left(\frac{Q_c^2 }{\mu^{\star2}}\right)}
  +\mathcal{O}\left(\log^{-2}\left(\frac{Q^2}{\mu^{\star2}}\right)\right)\Bigg]=\lim_{\omega \to 0} \left(\omega \frac{\rmd I}{\rmd\omega}\right)_{Q^2=Q_c^2}\, ,
  \end{split}
\end{equation}

where we neglected the dependency in $a$ in the logarithmic correction in the NLO term, since it is easily seen that it only contributes at higher orders. Thus, different choices for the multiplicative factor of $Q^2$, at NLO accuracy, only give rise to higher order logarithmic corrections. This observation has to hold to all orders in perturbation theory, since when all terms in the series are resumed, the spectrum is independent of the choice made for the matching scale. Therefore, the expansions differing in the choice of the matching scale and truncated at some order, can only differ by higher order corrections. This fact also allows to reduce the number of independent coefficients $c_{i,j}$ to be computed.\par 
In summary, we have shown that not only one has to allow for a dependence on $\omega$ in the matching scale for the perturbative expansion to be meaningful, the \textit{natural} choice for this scale is $Q^2\sim Q_c^2 \equiv \sqrt{\omega \hat{q}}$, and other choices for the matching scale only differ by subleading factors (assuming one uses $Q^2\sim\sqrt{\omega}$, which is the only physically reasonable scaling law for this problem). \par
\begin{figure}[h!]
    \centering
    \includegraphics[scale=.5]{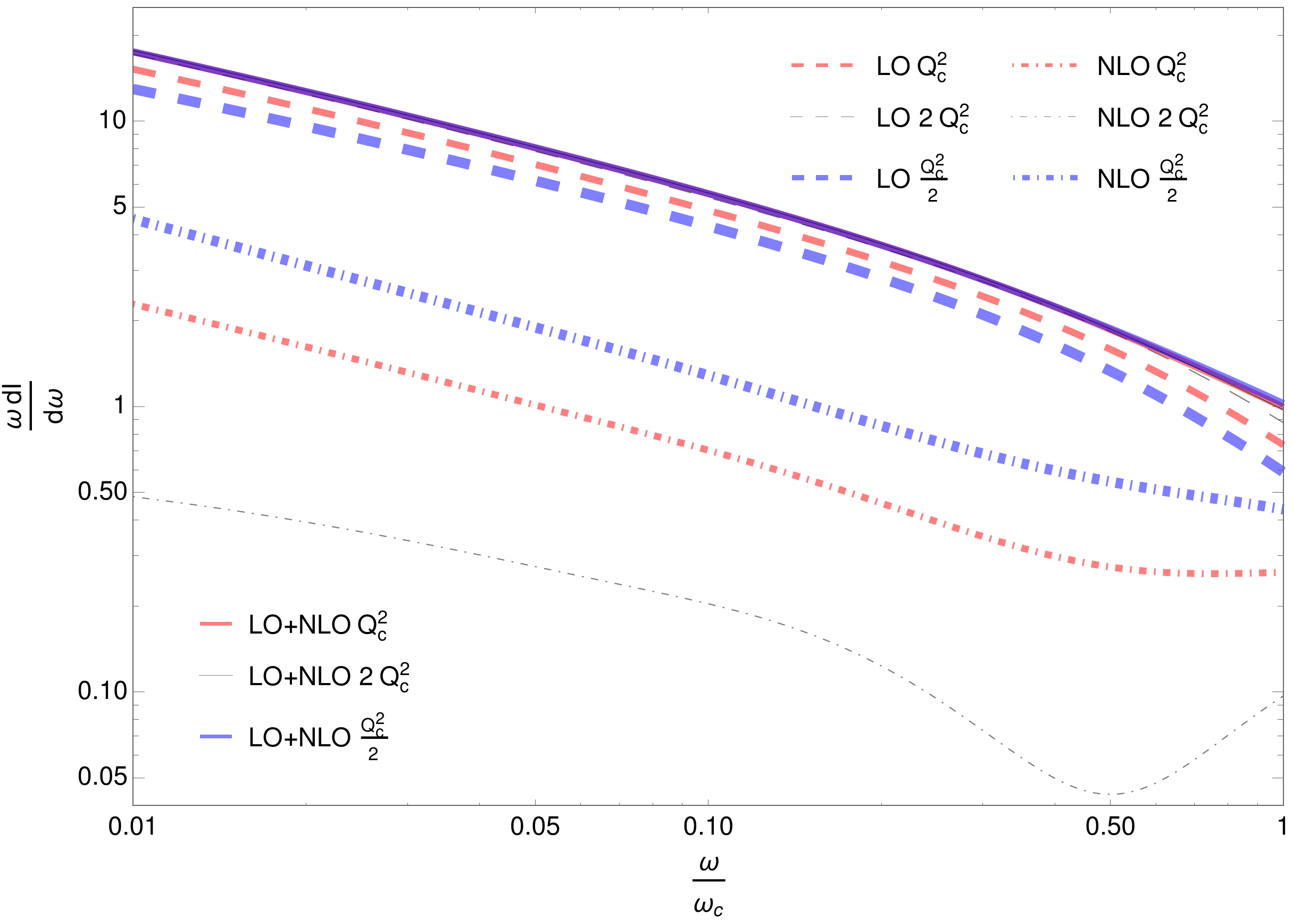}
    \caption{Calculation of the IOE at NLO accuracy, while fixing the matching scale $Q^2=Q_c^2=\sqrt{\hat{q}\omega}$ and varying this by $Q_c^2\to 2Q_c^2$ or $Q_c^2\to \frac{1}{2} Q_c^2$, where $\hat{q}=\hat{q}_0 \log\left(\frac{Q_0^2}{\mu^2}\right)$ and $Q_0^2=\hat{q}_0L$.}
    \label{fig:Q2_variation}
\end{figure}

 
In Figure \ref{fig:Q2_variation} we compute the spectrum at NLO accuracy while fixing the scale $Q^2\equiv Q_c^2$ (see figure for details) and then varying it by factors of $2$. We clearly see that the variation in the matching scale, in the low energy regime, lead to minimal modifications of the spectrum. In fact, we can see that this happens because while $Q^2$ increases the LO contribution increases but the NLO term becomes smaller, such that the contributions balance each other out, as shown in the analytical study. We would like to point out that this study is distinct from the one perform in \cite{Paper1} where one varied $\mu^{\star2}$. It is easy to see from the above expressions, that this does not lead to the same evolution as the one presented here.\par
Figure \ref{fig:Q2_variation} also shows that unlike the case where $Q^2$ is a fixed scale here, the spectrum does not diverge around $\omega \sim \hat{q}_0 \frac{L^2}{\log^2\left(\frac{Q^2}{\mu^{\star2}}\right)}$ and the LO and subleading terms balance each other out. This clearly shows that the \textit{interpolation problem} between the GLV and BDMPS-Z regimes requires a non trivial fix to how one defines the matching scale between the soft and hard regimes.\par 
Finally, it is also interesting to know how close to $\mu^{\star2}$ the matching scale $Q^2$, that decreases with $\omega$, can get, such that the NNLO is still significantly smaller than the LO and NLO terms. This translates into the sensitivity of the IOE to the approach to $\omega_{\rm BH}\sim \mu^{\star4}/\hat q$.  It is clear that when $Q^2\to \mu^{\star2}$ (or equivalently $\omega \to \omega_{\rm BH}$) every term in the expansion diverges. Starting from equation \eqref{eq:expansion_IOE_small_frequency_Q2_chosen}, we normalize the full spectrum in the low energy regime to the LO result and obtain
\begin{equation}
\lim_{\omega \to 0} \omega \frac{\rmd I}{\rmd\omega}_{\rm norm.}=1+  \left(\frac{0.508}{\beta}\right)+\left(\frac{0.029}{\beta^2}\right) \, ,
\end{equation}
where $\beta=\log\left(Q^2_c/\mu^{\star2}\right)=\frac{1}{2}\log (\omega/\omega_{\rm BH})$.
If we want to compare the contribution of the NNLO versus LO+NLO we just need to compute 
\begin{equation}\label{eq:scaling_final}
\frac{Q^2_c}{\mu^{\star2}}= \exp\left(-0.254+ 0.002\sqrt{16129+\frac{7256}{\alpha}}\right) \, ,
\end{equation}
where $\alpha\equiv\frac{\rm \rm NNLO}{\rm 1+\rm NLO}$ gives the percentile contribution of the NNLO term compared to the LO+NLO (up to NNNLO corrections). \par 
To proceed, we wish to discuss this scaling in terms of the Bethe-Heitler frequency $\omegaBH\equiv \mu^{\star4}/\hat{q}_0$ (recall $Q^2/\mu^{\star2}=\sqrt{\omega/\omegaBH}$). Then we can rewrite \eqref{eq:scaling_final} as 
\begin{equation}
\omega= \exp\left(-0.508 + 0.004\sqrt{16129+\frac{7256}{\alpha}}\right) \omegaBH\,,
\end{equation}
where we have chosen the positive root since it is the one of physical relevance.\par 
\label{eq:expansion_IOE_small_frequency_Q2_chosen}We then have that for $\alpha=1\%$, $\omega \geq 18.83 \ \omegaBH$; $\alpha=10\%$, $\omega \geq1.98 \ \omegaBH$  and for $\alpha=50\%$, $\omega\geq  1.21 \ \omegaBH$. The inequality symbol comes from the fact that the above equation gives the lower limit for $\omega$ below which the ratio $\rm \rm NNLO/ (\rm 1+\rm NLO)$ exceeds the value of $\alpha$. We see that the evolution with $\alpha$ is quite fast: when one requires $\alpha \sim 1\%$ the limit frequency has to be one order of magnitude larger than $\omegaBH$, but when $\alpha\sim 10 \%$ the limit frequency is of the order of $\omegaBH$. This shows that for the NNLO terms to be negligible (say giving less than $10 \%$ of the total contribution to the spectrum) compared to the LO and NLO terms, is not strongly dependent on low momentum tail and any typical energy scale would satisfy the inequalities presented above. Conversely, choosing matching scales which are essentially of the order of the Bethe-Heitler scale leads to the breakdown of the perturbative expansion, as expected (notice that when $\omega=\omegaBH$, equation \eqref{eq:expansion_IOE_small_frequency_Q2_chosen} becomes meaningless).

\section{Discussion and Outlook}\label{sec:Discussion_of_results}
In this paper we have provided an analytical and numerical study of the Improved Opacity Expansion at up to NNLO accuracy. In addition, we have presented, for the first time, a map between the GW and HTL models for the elastic in-medium cross section and the set of \textit{physical} parameters at leading logarithmic accuracy.  This results are best summarized in figure \ref{fig:GW_vs_HTL}, where it is clear that using our map with the GW full potential gives back an extremely good approximation of the HTL potential, up to dipole sizes $|\x|\sim 2/m_{\rm D}$. The combination of both these results, guarantees that we have a complete and systematic control over the analytic structure of the emission spectrum \eqref{eq:spectrum_general}.
This mapping is crucial since it gives meaning to comparisons between emission spectrums using different medium models. In the particular case of the IOE, we showed that this allowed us to have a full control over the accuracy of our result. \par
Moving on to a more detailed discussion of the IOE, our study allowed us to show that in the large frequency domain the spectrum is strongly dominated by the NLO term, which follows the well known GLV scaling. All other orders in the expansion, are power suppressed by factor of $\Bar{\omega}_c/\omega$. In particular we showed that the LO and NNLO terms are of the same order. However, an all order closed form formula is not possible to write down since, as argued above, as one moves away from the strict high energy limit, new contributions appear which are not power suppressed. \par 
On the opposite end of the spectrum, we found that the IOE has an extremely rich and interesting structure. In this limit, the LO term is the dominant contribution to the expansion, but important logarithmic contributions appear, order by order. This is in opposition to the high frequency regime, where NLO term is dominant over power suppressed contributions. \par 
In order to better understand the structure of the IOE in the small frequency domain we first noticed that for a fixed matching scale the expansion is ill defined and this lead us to conclude that there exists a \textit{natural} scale $Q_c
^2=\sqrt{\hat{q}\omega}$ which guarantees that the $\omega$ dependency of the matching is such that mutual cancellation between the many orders of the IOE guarantee that the full spectrum is finite. In addition, we showed that rescalings of $Q_c^2$ only affect higher order terms in the IOE (see \eqref{eq:expansion_around_QS}). This was numerically confirmed by the results in figure \ref{fig:Q2_variation}. Additionally, we want to point out that the exercise shown in figure \ref{fig:Q2_variation} clearly demonstrates that the interpolation between the GLV and BDMPS-Z regimes requires a proper treatment as the one provided by the IOE, and does not allow for a simplistic interpolating procedure. In fact, we have shown that the correct contribution to the spectrum in the region $\omega<\omega_c$ needs both the LO and the NLO terms in order to describe the correct result\par
Both these results are a direct consequence of the fact that the spectrum's dependence on matching scale must vanish when all orders in the IOE are taken into account. In fact, this observation means that after all terms are taken into account the spectrum must be of the form (for a general $Q^2$ scale;  see equation \eqref{eq:expansion_IOE_small_frequency})
\begin{equation}\label{eq:finalref}
  \lim_{\omega \to 0} \omega \frac{\rmd I}{\rmd\omega\rmd L}=\Bar{\alpha} \sqrt{\frac{\hat{q}_0 \, W\left(\frac{\sqrt{\omega \hat{q}_0}}{\mu^{\star2}}\right)  } {\omega}}\,,
\end{equation}
where $W$ is a general (unknown) function (introduced in \eqref{eq:expansion_IOE_small_frequency}), which captures all the finite corrections to the spectrum. Notice that the dependency in $Q^2$ disappears. From equation \eqref{eq:expansion_IOE_small_frequency_Q2_chosen} we can construct the W function order by order as
\begin{equation}
W^{\frac{1}{2}}\left(\frac{\sqrt{\omega \hat{q}_0}}{\mu^{\star2}}\right)=\log^{\frac{1}{2}}\left(\frac{\sqrt{\omega\hat{q}_0}}{\mu^{\star2}}\right)+\frac{0.508}{\log^{\frac{1}{2}}\left(\frac{\sqrt{\omega\hat{q}_0}}{\mu^{\star2}}\right)}+\frac{0.029}{\log^\frac{3}{2}\left(\frac{\sqrt{\omega\hat{q}_0}}{\mu^{\star2}}\right)}+\cdots \, ,
\end{equation}
where we have chosen the scale $Q^2=Q_c^2$.\par

It is straightforward to obtain to corresponding expansion of $W$, 
\begin{equation}
W\left(\frac{\sqrt{\omega \hat{q}_0}}{\mu^{\star2}}\right)=\log^{\frac{1}{2}}\left(\frac{\sqrt{\omega\hat{q}_0}}{\mu^{\star2}}\right)+\frac{1.016}{\log^{\frac{1}{2}}\left(\frac{\sqrt{\omega\hat{q}_0}}{\mu^{\star2}}\right)}+\frac{0.316}{\log^\frac{3}{2}\left(\frac{\sqrt{\omega\hat{q}_0}}{\mu^{\star2}}\right)}+\cdots \, ,
\end{equation}
These results show that the IOE admits to be written in a simple closed form for a fixed accuracy level with an additional prescription for the matching scale. All the results are valid so long as  the matching scale is chosen sufficiently higher than the Bethe-Heitler scale $\omegaBH$.\par 
Before moving on, we wish to point out that in \eqref{eq:finalref}, although the leading logarithmic behavior between each order truncation is well under control, there are logarithmic contributions order by order which might spoil the behavior of the series. Recall from above, we first showed that to have a proper converging series one has to require the matching scale to evolve with $\omega$ and then we showed that there is a \textit{natural} choice for this scale, with other choices (with the same scaling) varying only by subleading terms. However, before we ignored that when varying the scale $Q_c^2$ subleading terms (like $\log\left(\log\left(\frac{Q_c^2}{\mu^2}\right)\right)$) can be subleading in the number of logs but be of the order $\mathcal{O}(1)$\footnote{For instance, from equation \eqref{eq:expansion_IOE_small_frequency}, when expanding the LO term we obtain the leading contribution $\sim \log\left(\sqrt{\frac{\omega}{\omegaBH}}\right)$, while the subleading term reads $\sim \log\left(\log\left(\sqrt{\frac{\omega}{\omegaBH}}\right)\right)$. Therefore, normalizing to the LO term, the subleading term can contribute at NLO order (i.e. when counting the denominator logarithms) and can be an important factor since $\log\log$ might be of order of the leading coefficient $c_{1,0}$. This discussion follows to all orders and is a direct consequence of the fact that the matching scale is defined by a recursive equation.}. This is a direct consequence of the fact that the matching scale is  given by a recursive equation and one has to expand the recursive equation to a certain degree of accuracy. Then in the regime $\omegaBH\ll \omega\ll \omega_c$, the full spectrum should read 
\begin{equation}\label{eq:full_system}
\begin{split}
& \omega \frac{\rmd I}{\rmd\omega\rmd L}(\hat{q})=\omega \frac{\rm dI^{\rm LO}}{\rmd \omega\rmd L}(\hat{q}_{\rm eff})= \abar \sqrt{\frac{\hat q _{\rm eff}}{\omega}} \,,
\\& \hat{q}_{\rm eff} \equiv \hat{q}_0 W(Q^2_c/\mu^{\star2})= \hat{q}_0 \log\left(\frac{Q_c^2}{\mu^{\star2}}\right)\left[1+\frac{1.016}{\log\left(\frac{Q_c^2}{\mu^{\star2}}\right)}+\frac{0.316}{\log^2\left(\frac{Q_c^2}{\mu^{\star2}}\right)} +{\cal O}\left(\log^{-3}\left(\frac{Q_c^2}{\mu^{\star2}}\right)\right)\right]\,, 
\\& Q_c^2=\sqrt{\omega \hat{q}_0\log\left(\frac{Q_c^2}{\mu^{\star2}}\right)}\,.
\end{split}
\end{equation}
For example, at NLO accuracy one should use the NLO truncation of the second equation in \eqref{eq:full_system} and then use the second order expansion of the recursive equation for $Q_c^2$\footnote{Notice that this truncation includes the first double logarithmic contribution, as discussed in the previous footnote.}, as was done in \cite{Paper1,Paper2}.\par 
Then we have the remarkable result that in the small frequency regime, the full spectrum is captured by the BDMPS-Z solution with a renormalization of $\hat{q}$. Notice, that the above discussion where $Q_c^2\sim \sqrt{\omega \hat{q}_0}$ still holds when comparing the different orders of the IOE, but they might fail due to $\log\log$ contributions coming from the definition of the matching scale. Again, this exercise explicitly shows that the definition of the matching scale between the GLV and BDMPS-Z is a non-trivial problem and it can not be simply fixed \textit{ad-hoc}. \par 
Another important point in the work presented in this paper, is the evidence that the contributions to the spectrum coming from the NNLO order correction are parametric and numerically small. This ensures that, for example, in phenomenological applications, the LO+NLO truncation is sufficient. We have thus shown that the IOE provides a complete, systematic and self-consistent interpolation procedure between the GLV and BDMPS-Z pictures and this results holds as long as $\omega \gg \omegaBH$\footnote{ Extrapolating to near the scale $\omegaBH$ has already been studied at NLO accuracy \cite{Paper2}, although many questions are still to be answered.}. \par
\begin{figure}[h!]
    \centering
    \includegraphics[scale=.8]{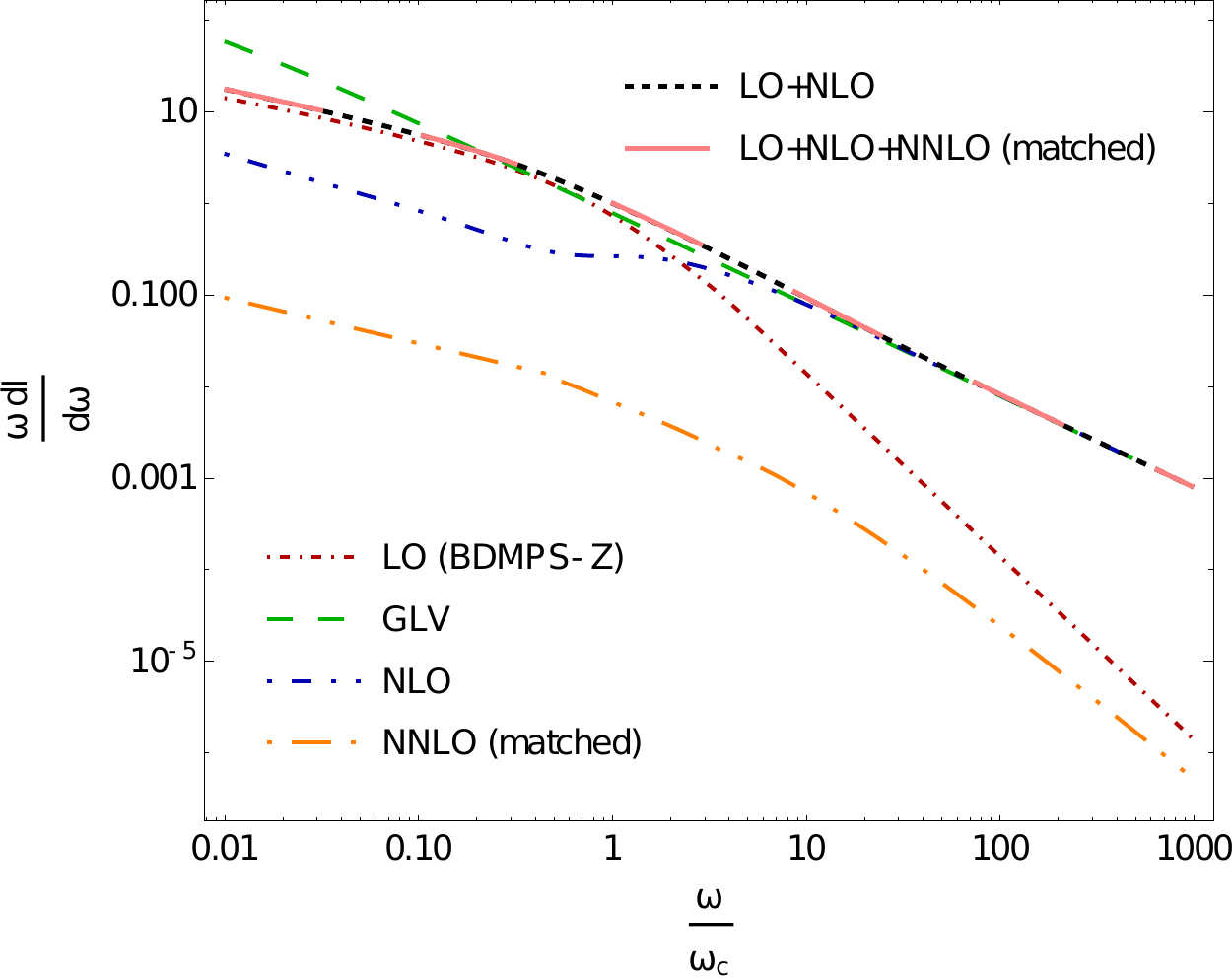}
    \caption{The analogous plot to figure \ref{fig:LO_GLV_NLO_NNLO}, but now extended to a larger frequency domain. In addition, we also include the spectrum up to NLO (black line) and NNLO (pink line) accuracy. All the curves are the same as in the previous plot, except the NNLO solution which is obtained by matching the scaling as $\omega \to 0$ at a frequency cut off $\omega_{cut}=0.5 \times \omega_c$, after which the solution is obtained by using equation \eqref{eq:NNLO_large_w}. The scaling for smaller frequencies is obtained by making use of equation \eqref{eq:w0_number} and the LO scaling law $\sqrt{\frac{\omega_c}{\omega}}$. This procedure is indicated by the \textit{matched} tag.}
    \label{fig:NNLO_match}
\end{figure}

Figure \ref{fig:NNLO_match} explicitly shows that including the NNLO term does not give any significant correction to the full spectrum. 
 In this plot we have extended the small frequency regime to large energies via the LO scaling at low energies: at intermediate $\omega$ the NNLO spectrum is obtained by extending the low energy result up to a matching scale, after which the high energy evolution at NNLO is used. We have tested this numerical procedure for several choices parameters $L$, $\mu$ and $\hat{q}_0$ and verified that, for reasonable matching scales $\omega_{match}\sim \omega_c$, the two ends of the spectrum nicely match each other. In addition, for several choices of parameters we have also seen that the NNLO contribution is always much smaller than the LO and NLO terms. \par
This work ensures that for future endeavours, it is sufficient to just keep track of the LO and NLO terms of the IOE. Therefore, taking into account all the results presented, in the future, we will be able to explore the single inclusive emission spectrum at NLO accuracy, while being able to have full control over the accuracy of the result. This is a key step for phenomenological implementations of the IOE.


\appendix 

\section{Useful integrals}\label{app:integrals}
In this appendix we shall calculate the following integral 
\begin{equation}
\int_0^\infty du \ \frac{u}{(u^2+b^2)\left(u^2+a^2\right)}\left(1-J_0(ux)\right) \,,
\end{equation}
that is related to the GW and HTL models by letting $b=a=\mu$ and $b=0$, $a=m_{\rm D}$, respectively.  
First we decompose the integrant as follows
\beq
&& \int_0^\infty du \ \frac{u}{(u^2+b^2)\left(u^2+a^2\right)}\left(1-J_0(ux)\right)=\nn
&& \frac{1}{(a^2-b^2)} \int_0^\infty du  \ \left[\frac{u}{(u^2+b^2)}-\frac{u}{ \left(u^2+a^2\right)}\right]\left(1-J_0(ux)\right)\,.
 \eeq
Now using the usual integrals  
\beq
  \int_0^\infty du  \ \left[\frac{u}{(u^2+a^2)}\right]J_0(xu) = K_0(a x) 
 \eeq
 and 
 \beq
 \int_0^\infty du \ \frac{u}{(u^2+b^2)\left(u^2+a^2\right)} = \frac{\log a^2-\log b^2}{2(a^2-b^2)}\,,
 \eeq
 we obtain 
\beq
&& \int_0^\infty du \ \frac{u}{(u^2+b^2)\left(u^2+a^2\right)}\left(1-J_0(ux)\right)=\frac{1}{(a^2-b^2)} \left[ K_0(ax) -K_0(bx)+\log a -\log b \right]\,. \nn
\eeq
There are two special cases that will correspond to the two models of interest. First, $a=b$  
 \beq
&& \int_0^\infty du \ \frac{u}{(u^2+a^2)^2}\left(1-J_0(ux)\right)=\frac{1}{2 a^2 } \left[ 1 -  a x  K_1(ax) \right]\,. \nn
\eeq
Then for $b=0$, using the form $K_0(bx) \approx - \log(b x/2) -\gamma_E$ 
\beq
&& \int_0^\infty du \ \frac{1}{u\left(u^2+a^2\right)}\left(1-J_0(ux)\right)=\frac{1}{a^2} \left[ K_0(ax) + \log(a x/2) +\gamma_E \right]\,. \nn
\eeq

\section*{Acknowledgements} 
We are grateful to Xabier Feal for helpful discussions.
This work is supported by the U.S. Department of Energy, Office of Science, Office of Nuclear Physics, under contract No. DE- SC0012704,
and in part by Laboratory Directed Research and Development (LDRD) funds from Brookhaven Science Associates. Y. M.-T. acknowledges support from the RHIC Physics Fellow Program of the RIKEN BNL Research Center.

The project that gave rise to these results received the support of a fellowship from ``la Caixa" Foundation (ID 100010434). The fellowship code is LCF/BQ/ DI18/11660057. This project has received funding from the European Union's Horizon 2020 research and innovation programme under the Marie Sklodowska-Curie grant agreement No. 713673. JB is supported by Ministerio de Ciencia e Innovacion of Spain under project FPA2017-83814-P; Unidad de Excelencia Maria de Maetzu under project MDM-2016-0692; European research Council project ERC-2018-ADG-835105 YoctoLHC; and Xunta de Galicia (Conselleria de Educacion) and FEDER. JB also acknowledges the support from the Fulbright Comission.

\bibliographystyle{elsarticle-num}

\bibliography{Lib.bib}


\end{document}